\documentclass{article}
\usepackage{arxiv}
\usepackage{graphicx,color,amsmath,amsfonts,amscd,amssymb,pstricks,tikz,cite,xcolor,import}
\usepackage{dcolumn}
\usepackage{bm}
\usepackage{longtable}
\usepackage{mathrsfs}
\usepackage{amsfonts}
\usepackage{textcomp}
\usepackage{esvect}
\usepackage{float}
\usepackage[USenglish,british]{babel}
\usepackage{environ}
\usepackage{xifthen}
\usepackage{xargs}			
\usepackage{hyperref}
\usepackage{physics}
\usepackage{multirow,here,tablefootnote}
\usepackage[separate-uncertainty=true]{siunitx}
\usepackage{physics,mathtools,comment}
\usepackage{hyperref}
\newcommand{\muf}{\mu_{\rm f} }
\newcommand{\tf}{t_{\rm fill} }
\newcommand{\tfo}{t_{{\rm fill},1} }
\newcommand{\tft}{t_{{\rm fill},2} }
\newcommand{\htfo}{\hat{t}_{{\rm fill},1} }
\newcommand{\htft}{\hat{t}_{{\rm fill},2} }

\newcommand{\tfj}{t_{{\rm fill},j} }
\newcommand{\htfj}{\hat{t}_{{\rm fill},j} }
\newcommand{\tffi}{t_{\rm fill,f} }
\newcommand{\htffi}{\hat{t}_{\rm fill,f} }

\newcommand{\htfop}{\hat{t}_{\rm fill} }

\newcommand{\tfc}{t_{\rm fill,c} }
\newcommand{\htc}{\hat{t}_{\rm c} }
\newcommand{\htfc}{\hat{t}_{\rm fill,c} }
\newcommand{\te}{t_{\rm e} }
\newcommand{\tre}{t_{\rm r}}

\newcommand{\tta}{t_{\rm ta} }
\newcommand{\ttai}{t_{{\rm ta},i} }
\newcommand{\ttaj}{t_{{\rm ta},j} }
\newcommand{\ttam}{t_{\rm ta,min} }
\newcommand{\ttao}{t_{\rm ta,1} }
\newcommand{\lt}{\mathcal{L}_{\rm tot} }
\newcommand{\li}{\mathcal{L}_{\rm int} }
\newcommand{\lmp}{L_{\rm mp} }
\newcommand{\lc}{\mathcal{L}_{\rm c} }
\newcommand{\lf}{\mathcal{L}_{\rm f} }
\newcommand{\lle}{\mathcal{L}_{\rm e} }
\newcommand{\nf}{N_{\rm fill} }
\newcommand{\nmp}{N_{\rm mp}}
\newcommand{\nin}{N_{\rm i}}
\newcommand{\Exp}[1]{\ensuremath{\mathrm{e}^{#1}}}
\newcommand{\eps}{\varepsilon}
\DeclarePairedDelimiter{\nint}\lfloor\rceil

\DeclareSIUnit\barn{b}
\DeclareSIUnit\proton{p}

\title{Optimisation of integrated luminosity in a circular collider with application to the LHC Run 2\thanks{This work has been supported by the CERN-INFN Project Associate programme}}
\author{F. Capoani\\
Dipartimento di Fisica e Astronomia, Universit\`a di Bologna and INFN Bologna, via Irnerio 46, Bologna, Italy \\
and \\
Experimental Physics Department CERN, Esplanade des Particules 1, 1211 Meyrin, Switzerland\\
\And
A. Bazzani\\
Dipartimento di Fisica e Astronomia, Universit\`a di Bologna and INFN Bologna, via Irnerio 46, Bologna, Italy\\
\And
B.~Giacobbe\\
INFN Sezione di Bologna, via Irnerio 46, Bologna, Italy \\
\And
M. Giovannozzi\thanks{Corresponding author: massimo.giovannozzi@cern.ch}\\
Beams Department, CERN, Esplanade des Particules 1, 1211 Meyrin, Switzerland}

\begin{document}
\maketitle

\begin{abstract}
Circular collider designs are tailored to maximise luminosity delivered to experimental detectors, effectively utilising the charged beams that have been accelerated for collisions. In reality, the key metric for the effective operation of a circular collider is the integrated luminosity provided to the experiments, which can significantly differ from the theoretical capability regarding instantaneous luminosity of the accelerator. Several factors influence the collection of integrated luminosity, with the most critical being the duration of each physics fill. This paper presents and examines strategies for determining optimal fill durations based on actual fill conditions, applying these methods to public luminosity data measured by the ATLAS detector during the LHC Run~2, covering the physics runs from 2016 to 2018.
\end{abstract}


%
\section{Introduction}
Modern circular colliders represent the pinnacle of technology, designed to maximise the luminosity provided to experimental detectors located in specially designated straight sections. The effectiveness of the design is validated by meeting the target values for key ring parameters, which require efficient beam commissioning to align the actual accelerator's performance with its design specifications. Moreover, efficient operation is crucial to achieve optimal performance, encompassing several different factors, some of which may be independent of the collider's performance, as these machines depend on a complex series of accelerators to supply the collider with high-quality beams. An often overlooked aspect in the operation of high-energy circular colliders is the effective management of the fill duration. The collider transitions between injection energy, where it is filled with particles, and top energy, when the counter-rotating beams are prepared and then brought into collision, maintaining this state for a specific period. Following this, the beams are disposed of and the collider's magnets revert to injection settings, readying for another fill. This procedure involves two distinct time scales: the fill duration and the turnaround time, the latter being the interval after the beam dump and before the next fill. The accumulation of integrated luminosity is significantly affected by the selection of these parameters, which, in turn, is influenced by typical beam parameters, hardware features, and fault rates across various accelerator systems. 

Optimisation of the operational cycle, which includes the determination of typical fill times, is part of the studies for the design of modern colliders, such as the upgrade of the LHC, the High-Luminosity LHC~\cite{BejarAlonso:2749422}, and has also been studied in the context of the Future Circular Collider study at CERN~\cite{FCC-eeCDR,FCC-hhCDR} where the optimisation of integrated luminosity has been considered in great detail~\cite{PhysRevSTAB.18.101002}. However, the perspective of these studies is radically different from that presented in this paper. In fact, the main interest is on the possibility of devising an approach that maximises the integrated luminosity during the actual run of a collider. This means that all aspects related to the variability of beam parameters and the operational stages, including the fill time and the turnaround times, are essential to this study that aims at incorporating at best all these effects.  

The CERN Large Hadron Collider (LHC)~\cite{LHCDR,evans:2012,myers:2013}, operating successfully at a beam energy of \SI{6.8}{TeV}, serves as an ideal environment to analyse the performance reach of a proposed optimisation strategy for the duration of a fill. This includes accounting for various factors related to the inherent variability of beam and hardware parameters. Recently, studies have been initiated to evaluate the potential increase in integrated luminosity through a refined strategy to optimise fill length~\cite{faletti}. The encouraging results have led to further research using advanced models for luminosity evolution~\cite{Giovannozzi:2018wmm,Giovannozzi:2018igq,amezza} and a detailed analysis of fault rates, which incorporates a crucial new factor into the optimisation strategy. Furthermore, a comprehensive analysis was conducted using Run~2 data, representing the period from 2016 to 2018. This paper presents an optimisation framework specifically tailored for the duration of a physics fill, thoroughly examined through Monte Carlo simulations to evaluate its theoretical performance. Following this, LHC Run~2 data are revisited, first to identify potential improvements in integrated luminosity, and subsequently to simulate the physics runs applying the proposed optimisation plan, to evaluate the feasible actual gain in integrated luminosity. 

This research is well-aligned with worldwide initiatives aimed at enhancing the efficiency of accelerator operation and exploitation (see, e.g., Ref.~\cite{seidel:ipac22-frplygd1} and references therein). These initiatives typically employ conventional control theory methods or prevalent machine learning techniques, which are excellent for handling the extensive data produced during accelerator operations and can be used to discover optimisation strategies for improving collider performance. 

The structure of the paper is outlined as follows: Section~\ref{sec:general} offers general remarks on optimising fill duration for physics operations in a circular collider. In Section~\ref{sec:lumi-opt}, we present and analyse the suggested strategies for determining the optimal fill duration. Section~\ref{sec:monte-carlo} discusses an extensive series of Monte Carlo simulations performed to validate these methods and to provide an estimate of their performance in absolute terms, i.e. looking at the performance reach of a large set of realisation of physics runs. The Monte Carlo simulations are performed as an online optimisation tool, i.e. they simulate a yearly run of a collider and optimise each fill sequentially. The analysis of data from LHC Run~2 is explored in Section~\ref{sec:application}, providing figures that illustrate the distributions of key parameters from 2016 to 2018, and evaluating the effect of optimisation techniques through numerical simulations similar to rerunning physics experiments. This section introduces optimisation techniques that take an online approach to adjusting the fill length, similar to the approach applied with Monte Carlo simulations. In addition, an \textit{a posteriori} optimisation method is explored, which helps to establish an estimate of the maximum possible increase in integrated luminosity achieved by optimising the fill length. Lastly, conclusions are made in Section~\ref{sec:conc}. In Appendix~\ref{sec:app}, the mathematical details of the derivation of the intensity evolution under the effect of dynamic aperture (DA), which is the heart of the luminosity modelling used in this paper, are presented for some functional forms of general interest for the transverse beam distribution. 
\section{General considerations on the optimisation of the integrated luminosity} \label{sec:general}
\begin{figure}[htb]
\centering
\includegraphics[width=\textwidth]{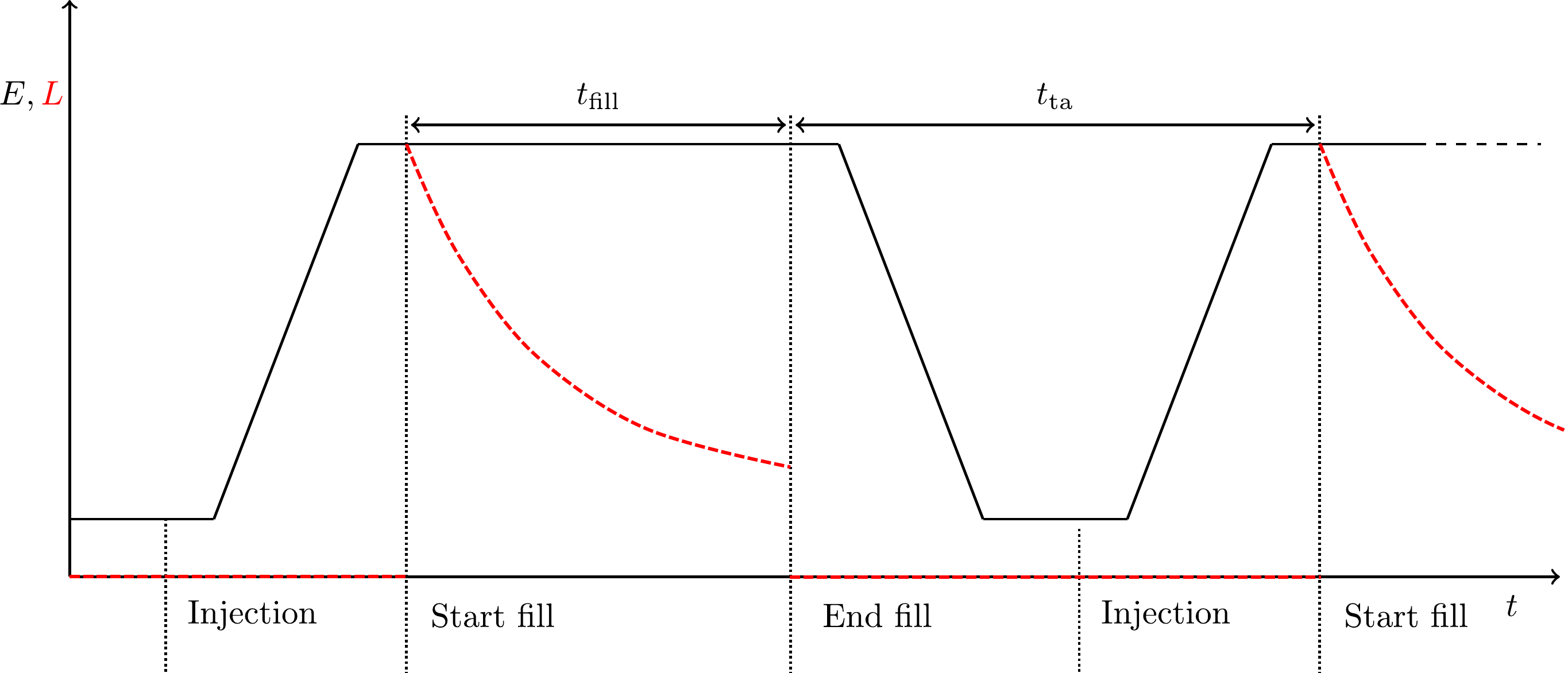}
\caption{Sketch of the luminosity production process in a circular collider. The black line represents the beam energy $E$ and the red dashed line the evolution of the luminosity $L$ as function of time $t$. $\tf$ represents the length of a physics fill while $\tta$ is the turnaround time.}
\label{fig:sketch}
\end{figure}

The process of luminosity production is sketched in Fig.~\ref{fig:sketch} where the magnetic cycle is shown and the main quantities are introduced.
In this study, $L(t)$ represents the evolution of luminosity during a fill. Any model can be considered, as the goal is to provide an optimised strategy to collect luminosity in a circular collider once a model is given to describe the function $L$.

The goal is to maximise the integrated luminosity and, in a purely deterministic case, this corresponds to maximising
\begin{equation}
    \begin{split}
        \lt(\tf) & = \nf \int_0^{\tf} \dd t \, L(t) \\
        & = \frac{T}{\tta+\tf} \int_0^{\tf} \dd t \, L(t) \, , 
    \end{split}
    \label{eq:main}
\end{equation}
where $\nf$ is the total number of fills during time $T$, the total time for physics, $\tta$ is the so-called turn around, i.e.\ the time elapsed from the end of a fill for physics and the start of the next one, $\tf$ is the length of the fill for physics. We note that the second step in Eq.~\eqref{eq:main} deserves some care. In principle, one should write
\begin{equation}
        \lt(\tf) = \nint[\Bigg]{\frac{T}{\tta+\tf}} \int_0^{\tf} \dd t \, L(t) \, , 
        \label{eq:main-spec}
\end{equation}
where $\nint{\cdot}$ stands for the nearest integer. However, under the assumption that $T \gg \tta +\tf$, which is always the case in reality, one can simply consider the form~\eqref{eq:main} that will be correct with a high degree of precision. 

Under these assumptions, the optimisation is performed noting that $\lt=\lt(\tf)$ and the maximum can be found by solving the equation
\begin{equation}
    \frac{\dd \lt}{\dd \tf} = 0 \, .
\end{equation}

Using the computations reported in Ref.~\cite{Giovannozzi:2018wmm}, it is possible to show that, for the case of pure burn-off and beam of equal intensity, the luminosity evolution is described by
\begin{equation}
L(t) = \frac{\Xi \, N^2_{\rm i}}{\left ( 1 + \sigma_{\rm int} \, n_{\rm c} \, \Xi \, \nin \, t \right )^2} \qquad \Xi = \frac{\gamma_{\rm r} f_{\rm rev}}{4 \, \pi \epsilon^* \beta^* \, k_{\rm b} } 
F(\theta_{\rm c}, \sigma_{z}, \sigma^* )
 \, ,
\label{eq:burnoff}
\end{equation}
where $\nin$ is the initial beam intensity, $n_\mathrm{c}$ is the number of collision points, $\sigma_\mathrm{int}$ represents the cross section of the proton-proton interaction, and the factor $F$ accounts for the reduction in volume overlap between the colliding bunches due to the presence of a crossing angle and is a function of the half crossing angle $\theta_{\rm c}$ and the transverse and longitudinal RMS dimensions $\sigma^*, \sigma_{z}$, respectively according to:
\begin{equation}
F(\theta_{\rm c}, \sigma_{z}, \sigma^* )=\frac{1}{\sqrt{1+
\left ( \displaystyle{\frac{\theta_{\rm c}}{2} \, \frac{\sigma_{z}}{\sigma^*}} 
\right )^2}} \, .
\label{geofac}
\end{equation}

Note that $\sigma^*=\sqrt{\beta^* \, \epsilon^*/(\beta_{\rm r} \, \gamma_{\rm r})}$, where $\beta_{\rm r}$ is the relativistic $\beta$-factor and $\epsilon^*$ is the value of the normalised RMS emittance.

Then, one obtains 
\begin{equation}
        \lt(\tf) = \frac{T}{\tta+\tf} \frac{\Xi \, N^2_{\rm i} \tf }{1+\sigma_{\rm int} \, n_{\rm c} \, \Xi \, \nin \, \tf} \, ,     
\end{equation}
and the optimal fill time is found as
\begin{equation}
    \htfop =  \sqrt{\frac{\tta}{\sigma_{\rm int} \, n_{\rm c} \, \Xi \, \nin}}\, ,
    \label{eq:opt-fill}
\end{equation}
where here and in the rest of the paper the hat is used to indicate the optimised value of the corresponding variable. 

The solution~\eqref{eq:opt-fill} indicates the dependence of the optimal fill time on some relevant parameters, namely $\tta$ and $\nin$, which is very useful for determining trends. However, these parameters are typically fluctuating from fill to fill, which reveals the key obstacle in applying this simple approach to operating a real collider. Although understanding both the theoretical and actual distributions of these parameters can help determine the distribution of the ideal fill length, it does not provide a means to define the duration of the current fill. This is because it depends on the previous and forthcoming fills, necessitating a comprehensive strategy to optimise the overall integrated luminosity. This is the topic of the strategy presented and discussed in the rest of this paper.

Before considering the general problem, one can consider the following intermediate case, namely, given $n$ values $\ttai$ distributed according to an exponential function and representing $n$ realisations of the turnaround time, one should maximise
\begin{equation}
    \lt(\tf)=n(\tf) \int_0^{\tf} \dd t \, L(t) \, ,
    \label{eq:opttest}
\end{equation}
which assumes that the fills should be of equal length although the turnaround times $\ttai$ are not, with the constraint
\begin{equation}
    \sum_{i=1}^n \ttai + n(\tf)\, \tf=T \, .
    \label{eq:constr}
\end{equation}

One can assume to replace the sum of the exponentially-distributed values of the turnaround time by a term $n \,  t_{\rm ave}$ so that 
\begin{equation}
    n(\tf)= \frac{T}{t_{\rm ave} + \tf} \, , 
\end{equation}
and the optimisation of Eq.~\eqref{eq:opttest} becomes the same type as the problem~\eqref{eq:main}. Note that implicitly one is assuming $t_{\rm ave} = \langle \ttai \rangle$ and the optimal fill time is rather an average optimal fill time.

It can be checked \textit{a posteriori} that the assumption of optimising the integrated luminosity by using equal fill lengths is the correct one. In fact, one could consider the situation in which a run is made of fills of the same length except for the last one. In this case, the integrated luminosity to be optimised reads
\begin{equation}
    \begin{split}
    \lt(\tfo, \tft) & =(n(\tfo,\tft)-1) \int_0^{\tfo} \dd t \, L(t) + \int_0^{\tft} \dd t \, L(t) \\ 
    & = \frac{T-(\tft+t_{\rm ave})}{t_{\rm ave} + \tfo}  \int_0^{\tfo} \dd t \, L(t) + \int_0^{\tft} \dd t \, L(t) \, ,
    \end{split}
    \label{eq:opttest1}
\end{equation}
where the constraint~\eqref{eq:constr} has been adapted to this new case. The maximisation of $\lt(\tfo,\tft)$ is obtained by solving
\begin{equation}
    \begin{cases}
        \displaystyle{\frac{\partial \lt}{\partial \tfo}} & = 0 \\
        & \\
        \displaystyle{\frac{\partial \lt}{\partial \tft}} & = 0  \, , 
    \end{cases}
\end{equation}
and by direct inspection, one finds that $\htfo=\htft$. 

\section{Optimisation of the integrated luminosity} \label{sec:lumi-opt}
\subsection{General approach}
The goal of determining the optimal fill time for the physics run of a circular collider can be addressed by defining a global function that should be optimised, and we propose to determine $\htc$, i.e.\ the optimal fill time of the current fill by maximising the following function
\begin{equation}
    \begin{split}
    \lt(\tfc,\tffi) & = \sum_{j=1}^{i-1} \li(\htfj) + \int_0^{\tfc} \dd t \, L(t) + \frac{T-\left [ \displaystyle{\sum_{j=1}^{i-1}} \left ( \ttaj+\htfj\right ) + \ttai + \tfc \right ]}{\frac{1}{i}\displaystyle{\sum_{j=1}^i} \ttaj+ \tffi} \int_0^{\tffi} \dd t \, \lmp(t) \, , \\
    & = \sum_{j=1}^{i-1} \li(\htfj) + \mathcal{L}_\text{c} (\tfc) + \mathcal{L}_\text{f} (\tfc,\tffi) \, , 
    \end{split}
    \label{eq:opt}
\end{equation}
where $\ttaj, \, 1 \leq j \leq i$ and $\htfj, \, 1 \leq j \leq i-1$ are the turnaround times and the optimal fill times, respectively, for all the fills from $1$ to $i$. Furthermore, the fill $i$ is the current one, whose integrated luminosity is indicated with $\lc$. Here, $\li$ stands for the integrated luminosity in a single fill. The optimised function also includes an estimate of the integrated luminosity collected in future fills, which is globally indicated as $\lf$ and built using estimates of future fills. In particular, $\lmp(t)$ stands for the most probable value of the function that represents the evolution of luminosity. It is clear that in case the luminosity function is Gaussian distributed, the most probable value corresponds to the average value. This function should be based on the knowledge derived from the past fills, if available. Note that an additional optimisation parameter has been introduced, namely $\tffi$, which represents the optimal fill time for all future fills. The third term in Eq.~\eqref{eq:opt} introduces a relationship between $\tfc$ and $\tffi$. Moreover, the term
\begin{equation}
    \frac{1}{i}\sum_{j=1}^i \ttaj
\end{equation}
is intended to provide an estimate of the average turnaround time.

The optimisation is performed as usual, namely
\begin{equation}
    \begin{cases}
        \displaystyle{\frac{\partial \lt}{\partial \tfc}} & = L(\tfc)- \displaystyle{\frac{1}{\frac{1}{i}\displaystyle{\sum_{j=1}^i} \ttaj+\tffi} \int_0^{\tffi} \dd t \, \lmp(t)} = 0 \\
        & \\
        \displaystyle{\frac{\partial \lt}{\partial \tffi}} & = \lmp(\tffi)- \displaystyle{\frac{1}{\frac{1}{i}\displaystyle{\sum_{j=1}^i} \ttaj+\tffi} \int_0^{\tffi} \dd t \, \lmp(t)} = 0  \, , 
    \end{cases}
\end{equation}
which gives
\begin{equation}
    \begin{cases}
        L(\htfc) & = \frac{1}{\displaystyle{\frac{1}{i}\sum_{j=1}^i} \ttaj+\htffi}  \displaystyle{\int_0^{\htffi} \dd t \, \lmp(t)} \\
        & \\
        \lmp(\htffi) & = \frac{1}{\displaystyle{\frac{1}{i}\sum_{j=1}^i} \ttaj+\htffi} \displaystyle{\int_0^{\htffi} \dd t \, \lmp(t)} \, ,
    \end{cases}
    \label{eq:lumi-opt-no-fail}
\end{equation}
where the second equation is used to determine $\htffi$ and the first $\htfc$. We note that as $L(t) \neq \lmp (t)$, and hence $\htfc \neq \htffi$. Therefore, a difference between the current and the future fills in terms of optimal fill time is expected. However, this situation occurs at the beginning of the physics run, as, asymptotically, $L(t) \to \lmp(t)$ and $\frac{1}{i}\sum_{j=1}^i \ttaj \to \left \langle \tta \right \rangle $ so that $\htfc - \htffi \to 0$. Another important point to note is that the solution does not depend on $T$. This is a consequence of the removal of the integer value function when defining the possible number of future fills in Eq.~\eqref{eq:opt}. This approximation is certainly irrelevant, unless one approaches the end of the run. 

We can solve Eq.~\eqref{eq:lumi-opt-no-fail} using the model of Eq.~\eqref{eq:burnoff} for the evolution of the instantaneous luminosity. We set $\varepsilon= \sigma_\text{int}n_c \Xi$, and
\begin{equation}
    L(t) = \frac{\alpha^2 \nmp^2}{(1+\varepsilon \alpha \nmp t)^2}\,,\qquad L_\text{mp}(t) = \frac{\nmp^2}{(1+\varepsilon \nmp t)^2}\,,
\end{equation}
with $\alpha=\nin/\nmp$ is the initial population of each fill relative to the most-probable value (which we assume to be known and fixed at $\nmp$); the value of $\varepsilon$ is assumed to be constant during a run. 

In that case, if $\langle \tta \rangle = \frac{1}{i}\sum_{j=1}^i \ttaj$, we obtain
\begin{equation}
    \htffi = \sqrt\frac{\langle\tta\rangle}{\nmp\varepsilon}\,, \qquad \htfc = \frac{\qty(1+\sqrt{\nmp\varepsilon \langle\tta\rangle})\alpha -1 }{\nmp\varepsilon\alpha}\, ,
    \label{eq:lumi-opt-simple-nofail}
\end{equation}
and
\begin{equation}
    \htfc = \htffi + \frac{1}{\varepsilon \nmp}\qty(1 - \frac{1}{\alpha}) \, ,
    \label{eq:htc-htf}
\end{equation}
which confirms that $\htfc -\htffi \to 0$ given that $\alpha \to 1$.

A plot of $\htfc$, obtained by solving Eq.~\ref{eq:lumi-opt-simple-nofail}, is shown in Fig.~\ref{fig:opt_simple} (top left). When $\nin/\nmp \leq 1/(1+\sqrt{\nmp\varepsilon\langle\tta\rangle})$ the optimal time is negative or zero and the optimisation shows a clear preference to immediately discard the fill to create a new one. All this depends on the square root of $\langle \tta \rangle$, indicating that the turnaround time is relevant, but the dependence is relatively mild.

The approach discussed above is certainly appropriate to describe the optimisation of the integrated luminosity when the remaining time is much longer than the typical run time. As the conclusion of the run nears, it is important to adopt an adjusted strategy that considers the remark regarding Eq.~\eqref{eq:main-spec}. The number of fills, while it should be part of the optimisation process, poses a challenge due to its integer nature, making it difficult to integrate into any optimisation framework. When the remaining time $\tre$ is comparable with that of two typical fills and a typical turnaround time, the function to be optimised is
\begin{equation}
    \lt(\tfc,\tre,\tta) = \int_0^{\tfc} \dd t\, L(t) + \int_0^{\tre-\tfc-\tta} \dd t\, \lmp(t) \, ,
\end{equation}
and the maximum is achieved by the solution of the equation
\begin{equation}
    L(\htfc)=\lmp(\tre-\htfc-\tta) \, , 
    \label{eq:last-two-fills0}
\end{equation}
which, for the simplified case of the simple model of luminosity evolution based on burn-off as in Eq.~\eqref{eq:burnoff}, is given by
\begin{equation}
    \htfc = \frac{\tre-\tta}{2} + \frac{1}{2 \varepsilon \nmp} \qty(1 - \frac{1}{\alpha}) \, .
    \label{eq:last-two-fills}
\end{equation}
Equation~\eqref{eq:last-two-fills} shows that the optimal fill time for the last two fills corresponds to keeping the two fills of equal length in the case where the luminosity of the current fill is similar to the most-probable case. Otherwise, a corrective factor applies. Of course, the value of $\tta$ is not known. Hence, one should determine the distribution of $\htfc$ given the distribution of $\tta$, evaluate the most-probable value and use it as $\htfc$. The integrated luminosity corresponding to this value of $\htfc$ should then be compared with that obtained by keeping the current fill for the entire lapse of time $\tre$ to check if a single fill would be more efficient.
\subsection{Including the failure probability of a fill}
The approach presented in the previous section does not include the observation that a long fill has a higher probability to fail than a short one. Hence, optimisation might propose an optimal fill time for which the risk of failure is high. Therefore, an efficient and realistic strategy to determine $\htfc$ should also include the effect of the probability of failure of a fill using a revised approach. 

Let $f(t)$ be the failure probability density, i.e.\ the probability that a fill will fail between $t$ and $t+\dd t$, and let $F(t)$ indicate the cumulative probability, i.e.\ the probability of a fill to fail before $t$. Consequently, $S(t)=1-F(t)$ represents the survival probability.

Given $t$ the time at which a fill should be terminated, the expected fill time can be expressed as
\begin{equation}
    \te(t) = \int_0^{t}\dd \tau \, \tau f(\tau) + t \int_{t}^{\infty} \dd \tau f(\tau) \, , 
\end{equation}
where we use a piecewise-defined function $\tilde t(\tau)$ to denote the effective time of a fill: if $\tau < t$, $\tilde t(\tau)=\tau$, and if $\tau \ge t$, $\tilde t(\tau) =t$, and we integrate $\tilde t(\tau)$ using the failure probability $f(\tau)$. Therefore, we obtain
\begin{equation}
\begin{split} 
\te(t) & = t F(t) - \int_0^t \dd \tau \, F(\tau) + t (1-F(t)) \\
& = t(1-S(t)) - \int_0^t \dd \tau \, (1-S(\tau)) + t S(t)\\
& = \int_0^t \dd \tau \, S(\tau ) \, .
\end{split}
\end{equation}

Similarly, the expected luminosity if we terminate the fill at $t$ becomes

\begin{equation}
    \begin{split}
        \lle(t) & = \int_0^{t}\dd \tau \, L(\tau) f(\tau) + L(t) \int_{t}^{\infty} \dd \tau f(\tau) \\
                & = \int_0^t \dd \tau \, L(\tau) S(\tau) \, .
    \end{split}
\end{equation}

This means that
\begin{equation}
    \lc(\tfc) =  \int_0^{\tfc} \dd \tau\, L(\tau) S(\tau)
\end{equation}
and the function to be optimised reads
\begin{equation}
    \lf(\tfc, \tffi) = \frac{T-\left [ \displaystyle{\sum_{j=1}^{i-1}} \left ( \ttaj+\htfj\right ) + \ttai + \te(\tfc) \right ]}{\displaystyle{\frac{1}{i}\sum_{j=1}^i \ttaj+ \te(\tffi)}} \int_0^{\tffi} \dd \tau\, \lmp(\tau) S(\tau) \, .
\end{equation} 

The optimisation of $\lc + \lf$ gives the following equations
\begin{equation}
    \begin{cases}
        L(\tfc) & =  \displaystyle{\frac{1}{\displaystyle{\frac{1}{i}\sum_{j=1}^i \ttaj} + \displaystyle{\int_0^{\tffi}} \dd \tau\, S(\tau) }} \displaystyle{\int_0^{\tffi}} \dd \tau\, \lmp(\tau) S(\tau) \, , \\
        \lmp(\tffi) & = \displaystyle{\frac{1}{\displaystyle{\frac{1}{i}\sum_{j=1}^i \ttaj} + \displaystyle{\int_0^{\tffi}} \dd \tau\, S(\tau)}} \displaystyle{\int_0^{\tffi}} \dd \tau\, \lmp(\tau) S(\tau) \, , 
    \end{cases}    
    \label{eq:lumi-opt-fail}
\end{equation}
which should be used to determine the unknowns $\htffi$ and $\tfc$: the first is calculated using the second equation, and the latter the first one.

\begin{figure*}[htb]
\includegraphics[width=.5\textwidth]{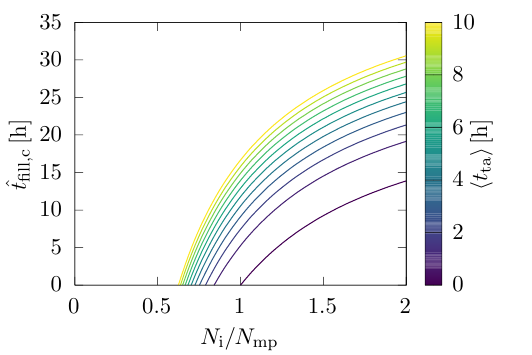}%
\includegraphics[width=.5\textwidth]{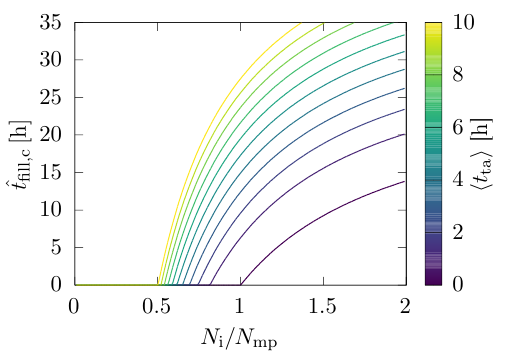} \\
\includegraphics[width=.5\textwidth]{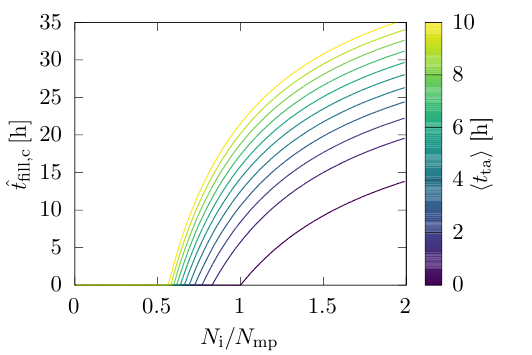}%
\includegraphics[width=.5\textwidth]{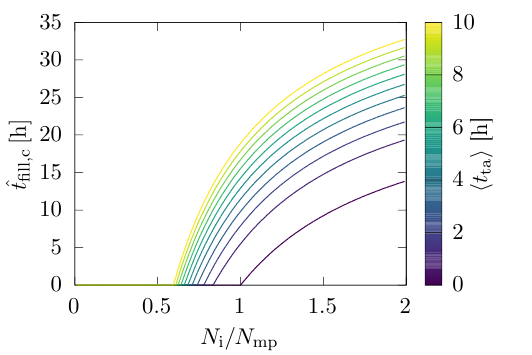}%
\caption{Value of $\htfc$ according to the model of Eq.~\eqref{eq:burnoff} as a function of the initial population of the fill relative to the most probable initial intensity. The top-left plot shows $\htfc$, when the possibility of fill failure is neglected, for different values of $\langle \tta \rangle$, having fixed $\varepsilon \nmp = \SI{1e-5}{\hertz}$. The subsequent plots introduce the survival probability $S(t)=e^{-\muf t}$ and show $\htfc$ for different values of $\mathrm{MTBF}=1/\muf$ (top right: $\SI{5}{\hour}$, bottom left: $\SI{10}{\hour}$, bottom right: $\SI{20}{\hour}$). Te top-left plot corresponds to a case in which $\mathrm{MTBF}=1/\muf \to +\infty$. Note that the LHC case (see Table~\ref{tab:datarun2}) is close to that represented in the bottom-right plot.}
\label{fig:opt_simple}
\end{figure*}

Applying the luminosity model of Eq.~\eqref{eq:burnoff}, and assuming an exponential failure distribution where $\muf$ is the inverse of the mean time between failures (MTBF), so that $S(t)=e^{-\muf t}$, we can numerically solve the equation for $\htffi$, and retrieve $\htfc$ from the equality $L(\htfc)=\lmp(\htffi)$, which gives
\begin{equation}
    \htfc = \htffi + \frac{1}{\varepsilon \nmp}\qty(1 - \frac{1}{\alpha}) \, ,
    \label{eq:htc-htf-MTBF}
\end{equation}
from which one observes that $\htfc - \htffi \to 0$ and that Eq.~\eqref{eq:htc-htf-MTBF} is equal to Eq.~\eqref{eq:htc-htf} and does not depend on $\muf$, making the relationship between $\htfc$ and $\htffi$ only dependent on the statistics that governs $L(t)$ and $\lmp(t)$. The results of the optimisation as a function of $\alpha$ and for four values of $\muf$ are shown in Fig.~\ref{fig:opt_simple} (bottom row and top right). Note that the top-left plot corresponds to a case in which $\mathrm{MTBF}=1/\muf \to +\infty$, and the convergence to this case is clearly seen. It is worth noting that the failure probability makes the optimisation more tolerant to fills with $\nin/\nmp < 1$, which are considered acceptable even when $\tta$ is short.

In the case of short turnaround time, the optimal fill time is largely independent of the fault rate and resembles the case where the fault rate is neglected. In contrast, larger values of the turnaround time have an impact on the optimal fill length value, smaller values of the MTBF inducing longer values of $\htfc$. 

Also in this framework, one should provide a strategy to deal with the last two fills of the run. In this case, when the remaining time $\tre$ is comparable with that of two typical fills and a typical turnaround time, then the function to be optimised is
\begin{equation}
    \lt(\tfc,\tre,\tta) = \int_0^{\tfc} \dd t\, L(t) S(t) + \int_0^{\tre-\tfc-\tta} \dd t\, \lmp(t) S(t) \, ,
\end{equation}
and the maximum is achieved by the solution of the equation
\begin{equation}
    L(\htfc) S(\htfc)= \lmp(\tre-\htfc-\tta) S(\tre-\htfc-\tta) \, , 
\end{equation}
or
\begin{equation}
    L(\htfc) = \lmp(\tre-\htfc-\tta) \exp \left [\muf \left (\tre-2\htfc-\tta \right ) \right ]\, .
    \label{eq:last-two-fills1}
\end{equation}

It is interesting to determine how the failure probability affects the optimal fill time, taking as reference the solution of Eq.~\eqref{eq:last-two-fills0}, always assuming the simplified case of the luminosity evolution based on burn-off as in Eq.~\eqref{eq:burnoff}. Indicating with $\Delta_\mathrm{c}$ the difference between the solution of Eq.~\eqref{eq:last-two-fills0} and that of Eq.~\eqref{eq:last-two-fills1}, and retaining only first-order terms in $\Delta_\mathrm{c}$, one obtains the following
\begin{equation}
    \Delta_\mathrm{c} = \frac{L \left ( \htfc \right ) \left \{ \exp\left [ \muf \left ( \tre -2 \htfc - \tta \right ) \right ] - 1 \right \}}{\dot{L}(\htfc) + \exp\left [ \muf \left ( \tre -2 \htfc - \tta \right ) \right ] \left (2 L \left ( \htfc \right ) + \dot{L}_\mathrm{mp} \left ( \tre - \htfc -\tta \right ) \right )} \, ,
\end{equation}
which shows that $\Delta_\mathrm{c} \to 0$ when $\muf \to 0$, and $\Delta_\mathrm{c} \propto \muf$. On the other hand, $\Delta_\mathrm{c} \to L \left ( \htfc \right )/\left (2 L \left ( \htfc \right ) + \dot{L}_\mathrm{mp} \left ( \tre - \htfc -\tta \right ) \right )$ when $\muf \to \infty$, and this limit is achieved exponentially. 
\section{Monte Carlo simulations of the proposed optimisation strategies} \label{sec:monte-carlo}
\subsection{General considerations on the Monte Carlo simulations}
The approach of integrated luminosity optimisation described in Section~\ref{sec:general} will be applied to a luminosity model more realistic than that of Eq.~\eqref{eq:burnoff}, which has the added value of being analytically solvable. To do that, we model the luminosity decay during a fill as a function of the turn number $\bar{t}$ following~\cite{Giovannozzi:2018wmm}, therefore including losses due to nonlinear dynamics in addition to the collision burn-off, i.e.
\begin{equation} \frac{L(\bar{t})}{L(0)} = \displaystyle{\frac{1}{\left [ 1 + \varepsilon \, \nin \, (\bar{t}-1) \right ]^2} -\left [ 1 + D^2(\bar{t}) \right ] e^{-D^2(\bar{t})} } \left \{ 2- \left [ 1 + D^2(\bar{t}) \right ] e^{-{D^2(\bar{t})}} \right \} \,,
\label{eq:lumimodel}
\end{equation}
where
\begin{equation}
    D(\bar{t}) =  \rho_\ast \left ( \frac{\kappa}{2e} \right )^\kappa \, \frac{1}{ \ln^\kappa \bar{t}} 
\end{equation} 
represents the so-called DA (see, e.g.~\cite{PhysRevE.53.4067,PhysRevE.57.3432,Bazzani:2019csk} and references therein), which is the extent of the phase-space region where the particle motion remains bounded over time. In this context, the DA provides the impact of nonlinear effects on the beam dynamics and depends on the two parameters $\rho$ and $\kappa$. The first parameter is related to the strength of the nonlinearities present in the accelerator and $\kappa$ to the dimensionality of the phase space of the dynamics of charged particles in the accelerator.

First of all, a Monte Carlo generator for luminosity fills has been prepared. To this end, we have to randomly choose the fill properties in a plausible way by generating sets of $(\nin,\, \rho_\ast,\, \kappa,\, \tta,\, t_\text{fail})$ according to the observed statistics of those parameters. These parameters are enough to generate the evolution of luminosity according to Eq.~\eqref{eq:lumimodel}. We used a selection of the LHC Run~2 fills data to determine a constant value of $\kappa$ and then generated values of $\rho_\ast$ and $\nin$ according to a bivariate normal distribution using their observed averages $\langle \rho_\ast \rangle$ and $\langle \nin \rangle$, standard deviations $\sigma_{\rho_\ast}$ and $\sigma_{\nin}$ and correlation coefficient $\mathrm{Corr}(\rho_\ast, \nin)$ during each year, to simulate fills with the same parameter statistics as those observed during Run~2. Note that the details of the fill selection and the fit performed are presented and discussed in Section~\ref{sec:application}.

For what concerns the turnaround time, it was observed that those times are distributed according to an exponential distribution with a constant offset, corresponding to a minimum turnaround time that is related with the characteristics of the power converters and magnets of the LHC ring, as well as the features of the whole accelerator chain, i.e.\ 
\begin{equation}
   \tta = \ttam + \ttao, \qquad \rho(\ttao) = \mu_\text{ta} \, e^{-\mu_\text{ta} \, \ttao} \, ,
\end{equation}
where we introduced the parameter $\ttam$, which encodes the minimum fixed turnaround time, and $\ttao$, which is the variable contribution to the turnaround time, whose exponential distribution is defined by $\mu_\text{ta}$.

The information about the failure probability of a fill can be extracted from the observed failure times, which well agree with a Weibull distribution~\cite{HODGES199435,OHRING1998} with $k<1$, namely
\begin{equation}
    \rho(t_\text{fail}) = k \, \muf (\muf \, t_\text{fail})^{k-1} e^{-(\muf \, t_\text{fail})^k} \, .
    \label{eq:weibull}
\end{equation}

The values of $\ttam$, $\mu_\text{ta}$, $k$ and $\muf$ are estimated by observing the turnaround and failure statistics for the LHC Run~2. Specifically, the turnaround parameters are evaluated from the statistical momenta of the observed distribution, with $\mu_\text{ta}=\sigma(\tta)$ and $\ttam=\langle\tta\rangle-\mu_\text{ta}$, while the failure parameters are estimated by fitting the histograms with the Weibull distribution. The numerical values obtained from the analysis of the Run 2 fills and used for the Monte Carlo simulations are shown in Table~\ref{tab:datarun2} (see also Ref.~\cite{Run2cern} for additional details on the LHC Run~2).

\subsection{Results of Monte Carlo simulations}

The actual test of the optimisation strategy of the fill duration is performed by generating a pool of fills according to the statistics of the parameters observed in the years 2016, 2017 or 2018, and compare the total integrated luminosity over the total time for physics of a given year. This is accomplished using two different strategies to decide when to stop a fill for physics, after which, in a real run, the operators proceed with the ramp down of the magnets preparing the whole ring for injecting the next fill for physics. In the baseline, unoptimised case, we choose to terminate a fill after a random time, chosen according to a normal distribution that reflects the actual cut times of a given year. In the second case, we use the proposed approach to select the time to stop a fill as a function of its parameters (assuming knowledge of the decay parameters $\kappa$ and $\rho$ since the start of a fill). The same comparison is performed including or not the possibility of fills to fail. It should be noted that both the random failure times and the turnaround times are part of the set of fills for a given year, so that the performance of the proposed strategy can be assessed by head-to-head comparisons between unoptimised and optimised fills. Repetition of the same procedure for a large number of possible realisations of years of operation, usually \num{1e4}, enables an accurate statistical evaluation of the performance of the proposed optimisation strategy. All this is applied to three possible scenarios, whose parameters are set according to the statistics observed in each of the three years of Run~2. 

\begin{table}[htb]
    \caption{Parameters used for the three years of the LHC Run~2.}
    \centering
    \begin{tabular}{c|c|c|c|c}
              \hline
                   &      & \multicolumn{3}{c}{Year} \\
    Parameter      & Unit & 2016 & 2017 & 2018 \\
              \hline
    $\langle\rho_\ast\rangle \pm \sigma_{\rho_\ast}$  & & \num{201 \pm 14} & \num{351 \pm 10} & \num{232 \pm 9}\\ \hline
    $\kappa$     & & \num{.94} & \num{1.08} & \num{0.97}\\ \hline
    $\langle N_\text{i}\rangle \pm \sigma_{\nin}$ & \num{E14}~p & $\num{2.4(0.1)}$ & $\num{2.1(0.6)}$ & $\num{2.77(0.09)}$\\ \hline
    $\mathrm{Corr}(\rho_\ast, N_\text{i})$ & & \num{.327} & \num{.384} & \num{.025}\\ \hline
    $\epsilon^*$ & [\SI{}{\micro\metre}] & \multicolumn{3}{c}{\num{2.2}} \\ \hline
    $\beta^*$ & [\SI{}{m}] &\multicolumn{2}{c}{\num{.4}} & \num{.3} \\ \hline
    $\theta_\text{c}/2$  & [\SI{}{\micro\radian}] & \num{185} & \num{150} & \num{160} \\ \hline
    $\gamma_\text{r}$ & & \multicolumn{3}{c}{\num{6929.64}} \\ \hline
    $f_\text{rev}$  & [\SI{}{\kilo\hertz}] & \multicolumn{3}{c}{\num{11.245}} \\ \hline
    $n_\text{c}$  & & \multicolumn{3}{c}{\num{2}} \\ \hline
    $k_\text{b}$  & & \num{2220} & \multicolumn{2}{c}{\num{2556}} \\ \hline
    $\sigma_z$  & [\SI{}{m}]& \multicolumn{3}{c}{\num{.102}} \\ \hline
    $\sigma_\text{int}$  & [\SI{}{\milli\barn}] & \multicolumn{3}{c}{\num{81}} \\ \hline
    $\langle t_\text{fill}\rangle \pm \sigma_{t_\text{fill}}$ (actual) & [\SI{}{h}] & \num{18.5\pm6.9} & \num{11.5\pm3.6} & \num{11.5\pm3.7} \\ \hline
    $\ttam$ & [\SI{}{h}] & \num{2.56} & \num{1.79} & \num{1.9} \\ \hline
    $\ttao$ & [\SI{}{h}] & \num{5.86} & \num{4.20} & \num{4.26} \\ \hline
    $\mathrm{MTBF}=1/\muf$ & [\SI{}{h}] & \num{26.4\pm 3.6} & \num{19.3\pm 4.1} & \num{20.2\pm 2.6} \\ \hline
    $k$ & & \num{0.79\pm 0.08} & \num{0.70\pm 0.09} & \num{0.87\pm 0.09} \\ \hline
    Total time for physics~\footnotemark[1] & [\SI{}{h}] & \num{3504} & \num{3360} & \num{3480} \\ \hline
    \end{tabular}
    \label{tab:datarun2}
\end{table}
\footnotetext[1]{This is the total official duration of the proton-proton physics run.}

\begin{figure}[htb]
\centering
2016\\
\includegraphics[width=.31\textwidth]{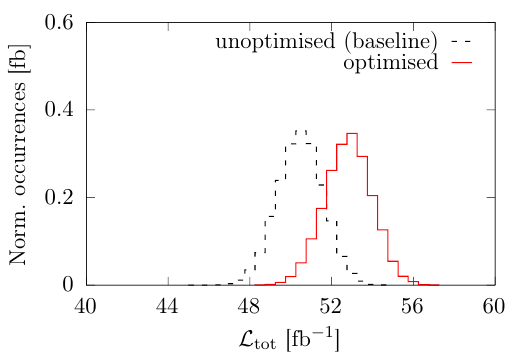}
\includegraphics[width=.31\textwidth]{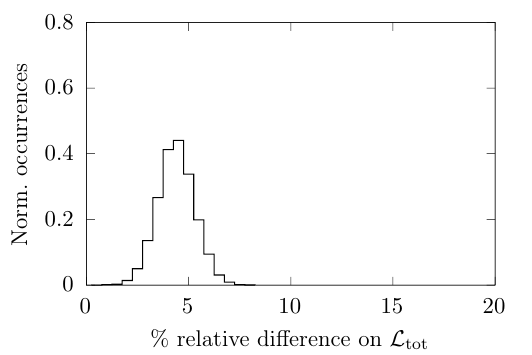}
\includegraphics[width=.31\textwidth]{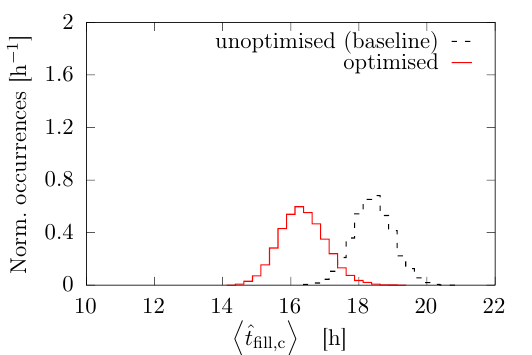}\\
2017\\
\includegraphics[width=.31\textwidth]{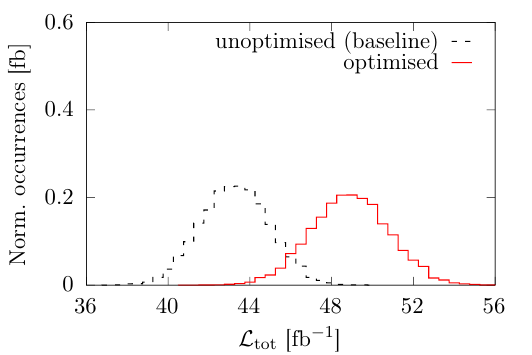}
\includegraphics[width=.31\textwidth]{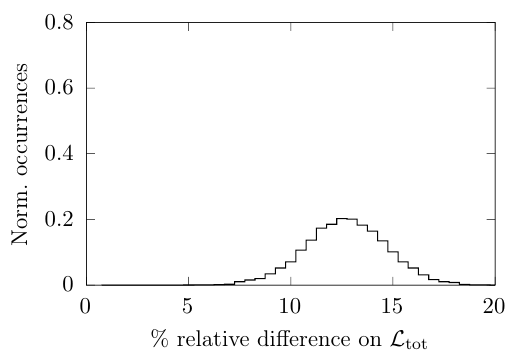}
\includegraphics[width=.31\textwidth]{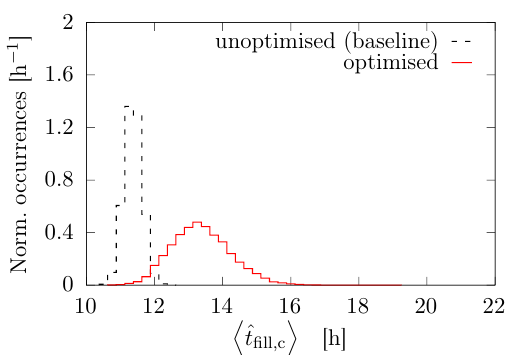}\\
2018\\
\includegraphics[width=.31\textwidth]{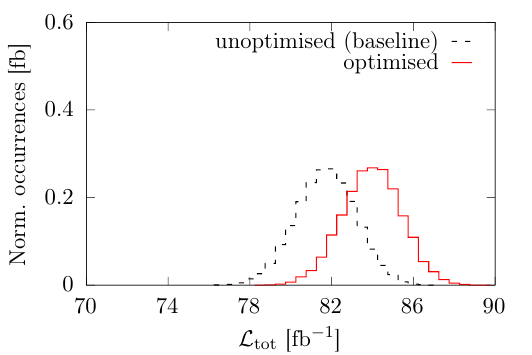}
\includegraphics[width=.31\textwidth]{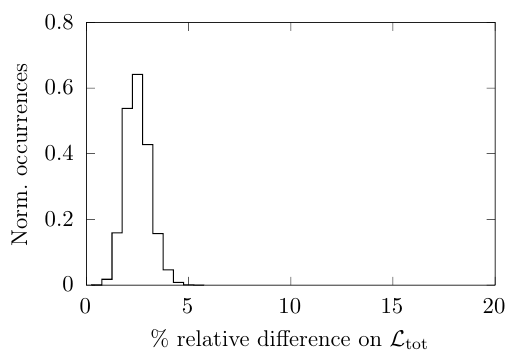}
\includegraphics[width=.31\textwidth]{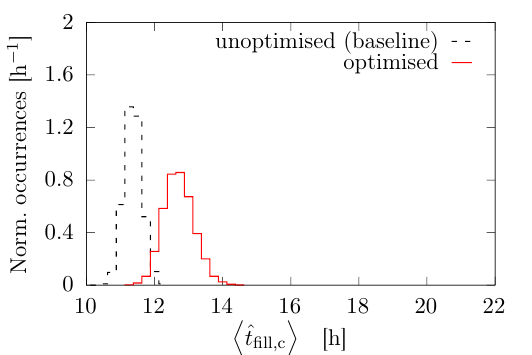}\\
\caption{Histograms of the total integrated luminosity (left column), the relative difference in integrated luminosity (centre column) and average fill time (right column) of \num{1e4} realisations of a year of LHC run. The parameters used are those of Table~\ref{tab:datarun2} for 2016 (first row), 2017 (centre row), and 2018 (bottom row) when a random fill time baseline (black) or the proposed optimisation approach (red) are used. No failure probability is included in these Monte Carlo simulations.}
\label{fig:online}
\end{figure}

\begin{figure}[htb]
\centering
2016\\
\includegraphics[width=.31\textwidth]
{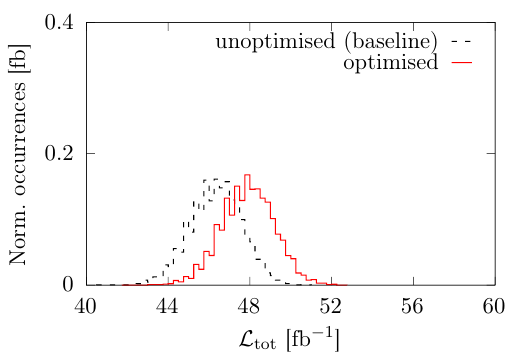}
\includegraphics[width=.31\textwidth]{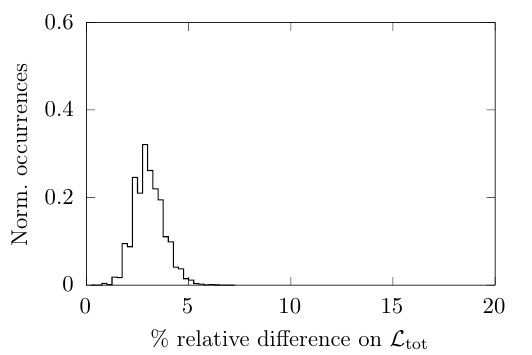}
\includegraphics[width=.31\textwidth]{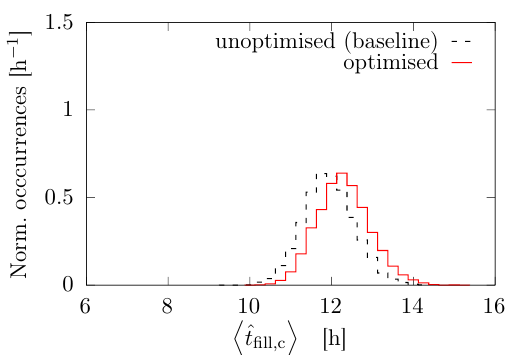}\\
2017\\
\includegraphics[width=.31\textwidth]{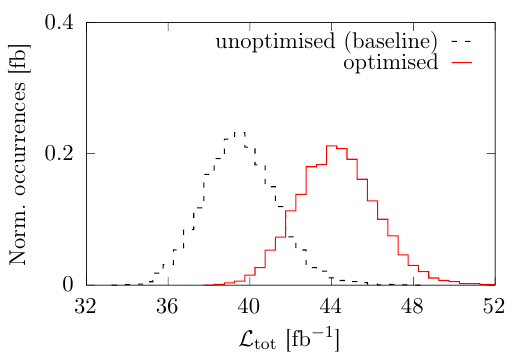}
\includegraphics[width=.31\textwidth]{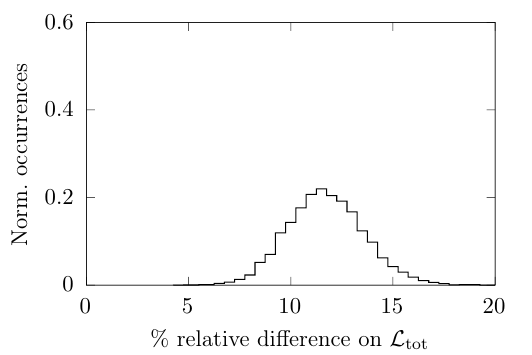}
\includegraphics[width=.31\textwidth]{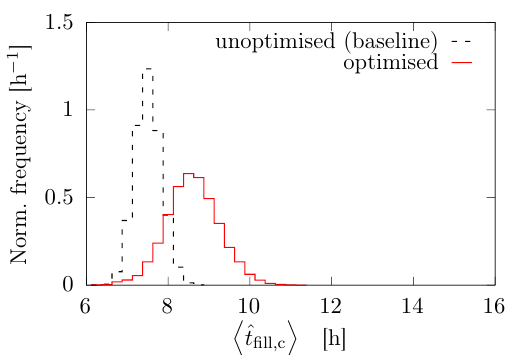}\\
2018\\
\includegraphics[width=.31\textwidth]{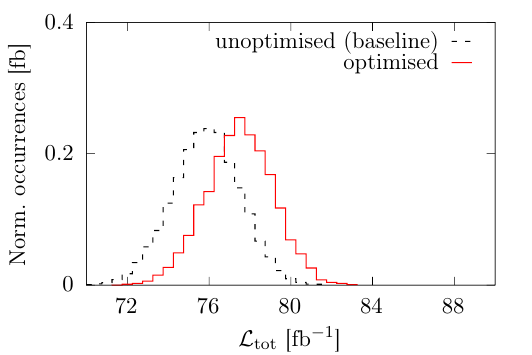}
\includegraphics[width=.31\textwidth]{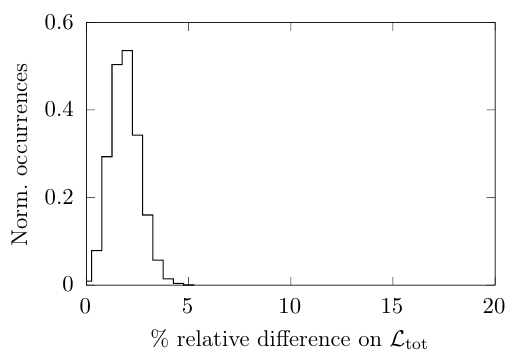}
\includegraphics[width=.31\textwidth]{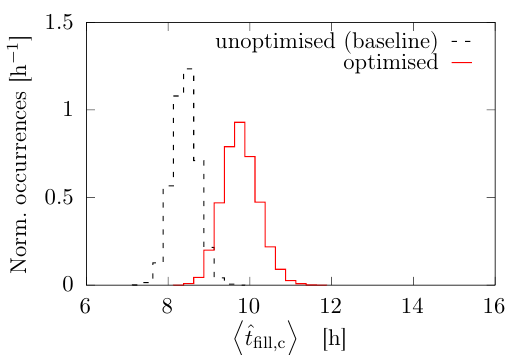}\\

\caption{Histograms of the total integrated luminosity (left column), the relative difference in integrated luminosity (centre column) and average fill time (right column) of \num{1e4} realisations of a year of LHC run. The parameters used are those of Table~\ref{tab:datarun2} for 2016 (first row), 2017 (centre row), and 2018 (bottom row) when a random fill time baseline (black) or the proposed optimisation approach (red) are used. Failure probability according to Eq.~\eqref{eq:weibull} is included in these Monte Carlo simulations, using the parameters listed in Table~\ref{tab:datarun2}.}
\label{fig:online_failure}
\end{figure}

Figures~\ref{fig:online} and~\ref{fig:online_failure} show histograms representing the distribution over the realisations of the yearly runs of the integrated luminosities with and without optimisation of the fill time, the relative difference in integrated luminosity between the two strategies and the distribution of the average fill time for the two approaches. In Fig.~\ref{fig:online} no failure is considered, while failure effects are included in the simulations whose results are shown in Fig.~\ref{fig:online_failure}. 

The first crucial observation is that the proposed optimisation strategy successfully achieves its goal of providing higher values of the integrated luminosity in each of the three years considered. The gain is clearly observed by looking at the distribution of the integrated luminosities (first column of Figs.~\ref{fig:online} and~\ref{fig:online_failure}), regardless of whether the possibility of failure is considered.  This is very clearly visible in the centre column of the same figures, where the distribution of relative difference in integrated luminosity is shown. These distributions are always in the positive domain of relative differences, indicating that the optimisation is efficient in the entirety of the case considered in the Monte Carlo simulations. The actual gain depends on the year, as the beam parameters are different, and on the presence of not of failure: without failure effects, the relative gain is slightly higher than in the corresponding case with failure effects included. 

Furthermore, we observe that, depending on the year and the presence or absence of failure effects, the optimisation process tends to distribute the optimal fill times around a mean value that can be either longer or shorter than the average fill length observed during the actual LHC operations. In some cases, fills are made, on average, shorter, thus increasing the time spent in turnaround operations. For instance, for the year 2016 data without failure, we observe that optimisation reduces, on average, the time spent in physics from 69\% to 66\% of the total run time of the year. However, this allows us to make room for more high-quality fills, thus increasing the total integrated luminosity by sensible amounts. 

\begin{figure}[htb]
\centering
\includegraphics[width=\textwidth]{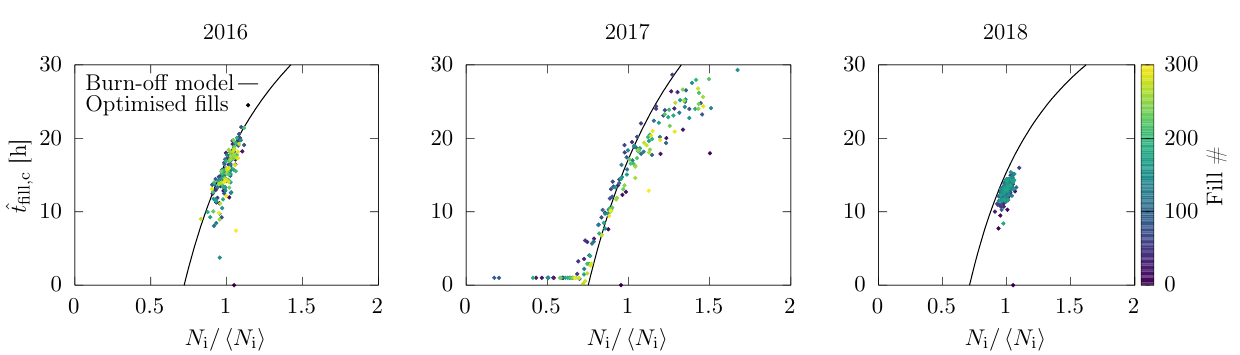}
\includegraphics[width=\textwidth]{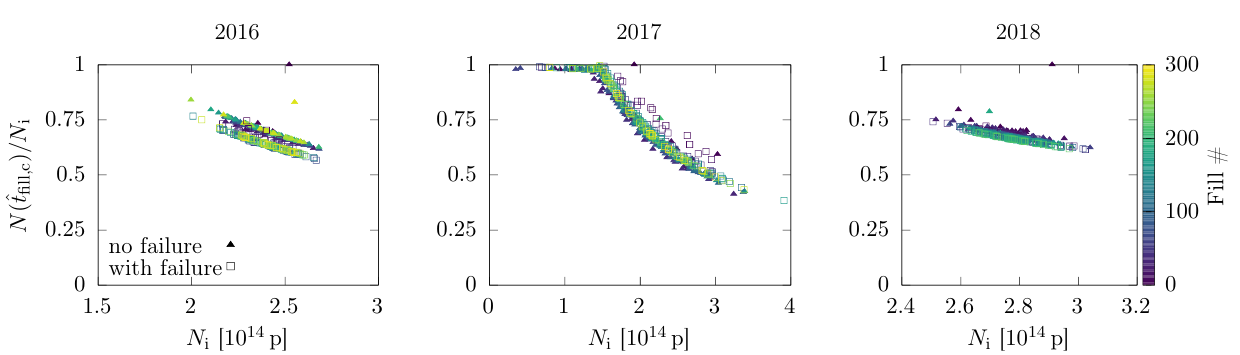}
\includegraphics[width=\textwidth]{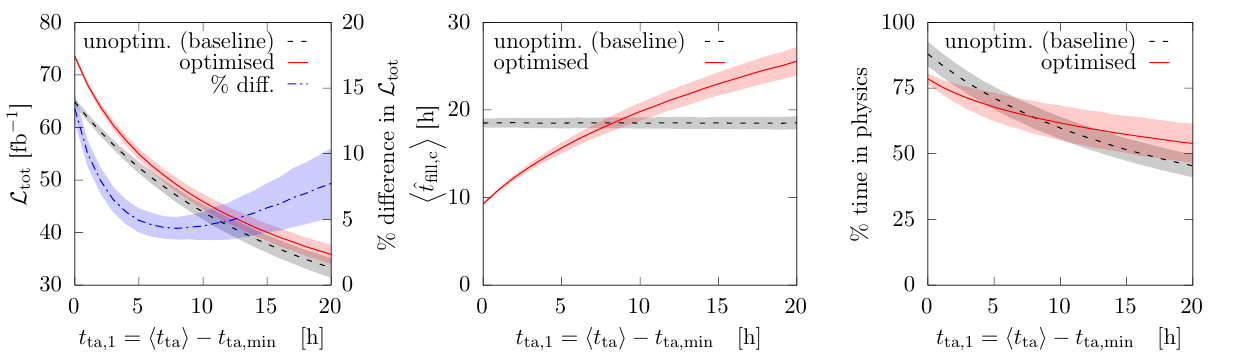}
\caption{Top row: optimised fill lengths for a single realisation of each year of Run~2 (using the parameters of Table~\ref{tab:datarun2}) as function of the initial beam population (normalised by its mean value) and comparison with the analytical value of the optimal time from the burn-off model of Eq.~\eqref{eq:htc-htf}. The colours identify the fills according to their ordinal position in the year's run. Centre row: ratio between the total beam intensity at the optimal cut time and the initial intensity, as function of the initial intensity for a single realisation of each year of Run~2 in the case without failure (full circles) and with failures (open squares). Bottom row: statistics on the optimised fills as a function of $\ttao$, describing the distribution of the turnaround time. The solid line represents the average, while the filled area represents one standard deviation around the average. The results are obtained using the 2016 parameters ($\ttam=\SI{2.56}{\hour}$). The left plot shows $\lt$ and the relative difference for the unoptimised (baseline) and the optimised scenarios, the centre plot $\htfc$, and the right plot the relative fraction of the total time spent in physics.}
\label{fig:online_compare}
\end{figure}

Additional analyses of the results of the Monte Carlo simulations are reported in Fig.~\ref{fig:online_compare}. In the top-row plots, we display the optimised times ($\htfc$) for a single realisation of an entire year's fill history as a function of the relative initial numerosity. The results of the Monte Carlo simulations are compared with the analytical estimate based on the solution of the burn-off model of Eq.~\ref{eq:htc-htf}. We note that the distribution of $\htfc$ tends to stay relatively close to the estimate provided by the analytical model. The few deviations are due to initial fills for which the estimate of the parameters needed to predict the luminosity of future fills, an essential element in the proposed optimisation strategy, is not reliable enough. However, the case of 2017 is more interesting, as it allows us to explore the behaviour of the optimisation algorithm when the beam parameters vary over a wider range. We observe that the optimisation process cuts to the minimum time (\SI{1}{hour}) set in the Monte Carlo, the fills with $\nin$ very close to the threshold value provided by the analytical model. On the other hand, when $\nin$ departs from the average value, the solution provided by the optimisation strategy moves away from the analytical solution: this is a clear sign that the nonlinear effects start to play a significant role, which results in a lower value $\htfc$ with respect to the analytical solution. 

In the centre row of Fig.~\ref{fig:online_compare}, it is shown, for a single realisation of each year, with and without failure effects, the relative value of the beam intensity at the optimised cut time, according to the formula
\begin{equation}
    \frac{N(\htfc)}{N_\text{i}} = \sqrt{\frac{L(\htfc)}{L_0}}\, .
\end{equation}

We observe that, regardless of whether failure effects are considered, most fills are cut when the intensity has reached approximately \SIrange{70}{75}{\%} if its initial value, corresponding to approximately \SI{50}{\%} of the initial luminosity: this could be a useful empirical rule, which, however, depends on the details of the physical parameters considered. Variations in this observation are observed to be evident in the data associated with the 2017 run. In this period, there was a consistent increase in beam intensity to enhance LHC performance. In these circumstances, the optimisation algorithm aims to significantly shorten the fills with beam intensity on the lower end of the distribution, thereby creating outliers.

Finally, in the bottom row of Fig.~\ref{fig:online_compare} shows an additional detailed analysis of the results of the Monte Carlo simulations applied, as an example, to the case of the 2016 run parameters. The focus is on the dependence of the optimisation results on the properties of the distribution of the turnaround time, as this might also be interesting for the general case of other circular colliders. A comparison is made between the case with unoptimised (baseline) and optimised management of the integrated luminosity for $\lt$ (left plot), $\htfc$ (centre plot) and the relative fraction of the total time spent in physics (right plot). The gain in $\lt$ provided by the optimisation is very large for short turnaround times, reaching a minimum and then increasing again for much longer turnaround times. Of course, the absolute value of $\lt$ steadily decreases with increasing turnaround time. We note that $\langle \htfc \rangle$ is a non-decreasing function of the turn around time, whereas the relative total time spent in physics is a decreasing function of the turnaround time. Finally, we remark that even when the average of $\htfc$ for the optimised approach happens to be equal to that obtained with the unoptimised (baseline) strategy, there is still an advantage in applying the optimisation strategy, as it adjusts $\htfc$ of each fill in dependence on the variations of the variable fill parameters, notably, $\nin$ and $\rho_\ast$.

The results of the Monte Carlo simulations are very promising, indicating that there is the possibility of increasing the integrated luminosity collected at the LHC, in the range 3\% to 11\% for the most-probable value as shown in the centre column of Figs.~\ref{fig:online} and~\ref{fig:online_failure}, when a well-defined strategy to adjust the length of a fill for physics is implemented. We stress here an essential aspect of these simulations: the implementation of the optimisation approach assumes perfect knowledge of the parameters of the fill. However, the fitting of real LHC fills on the model of Eq.~\eqref{eq:lumimodel}, which we will present in Section~\ref{sec:application}, might not achieve a reliable estimate of $\rho_\ast$ or $\kappa$ over a short time (typically \SI{1}{h}) of observation of the luminosity decay. This calls for the development of reliable and on-line modelling of the measured luminosity evolution, which is the focus of future research.

\section{Application to the case of the LHC Run~2} \label{sec:application}
\subsection{Performance analysis of the fit of the luminosity evolution}
To construct the integrated luminosity optimiser and apply it to LHC Run~2 data, we start from the publicly available ``Massi'' summary files~\cite{massifiles}, which are provided by the ATLAS collaboration~\cite{avoni:2018,ATLAS:2023}, reporting, for each fill of the LHC, the measured luminosity expressed in $\si{\hertz/\micro\barn}$ and collected by each of the four LHC experiments at a typical time interval of \SI{56}{\second}. All data used in these studies record the luminosity measured by the ATLAS experiment~\cite{atlas}.

In principle, the model of Eq.~\eqref{eq:lumimodel} could be fitted to the luminosity data to retrieve the characteristic model parameters for each year of Run~2. In reality, great care is needed. First of all, not all fills for which luminosity data exist correspond to successful physics fills. In this respect, we decide to retain in our dataset only the data of those fills that were not terminated by any sort of technical issue (the information about the fills terminated abnormally is used to extract the information about the failure effects) or which suffered technical issues during the turnaround time preceding them. Furthermore, we exclude fills with too low an initial intensity (less than $\SI{1.5e14}{\proton}$ for 2016 and 2018 and less than $\SI{1e14}{\proton}$ for 2017), as well as very short fills (less than $\SI{1}{\hour}$). These types of fill are typical of the so-called intensity ramp-up period and are neither representative of typical high-luminosity fills nor contribute significantly to the integrated luminosity; hence the decision to discard them. Note that during the intensity ramp-up period the number of bunches $n_\mathrm{b}$ is changed: the selection criterion on $\nin$ allows the retained fills to have $n_\mathrm{b}$ equal to or close to the nominal value for a given year. Finally, we also discard fills in which luminosity levelling is performed. This consists of changing the optical parameters of the LHC ring, $\beta^\ast$, transverse beam separation, and the crossing angle to keep the instantaneous luminosity constant during the burn-off of the protons due to collisions. This operational mode was tried occasionally in 2017 and more regularly in 2018, but, overall, represents a tiny fraction of the LHC fill set. The study of luminosity optimisation in the presence of luminosity levelling, which is currently used during Run~3 and is the default for the LHC luminosity upgrade~\cite{hl-lhc-tdr}, is left for future analysis.

\begin{figure}
\centering
\includegraphics[width=.4\textwidth]{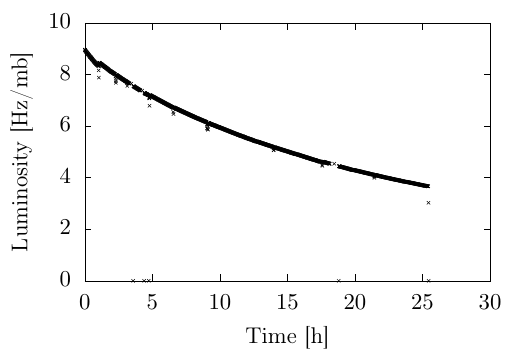}
\includegraphics[width=.4\textwidth]{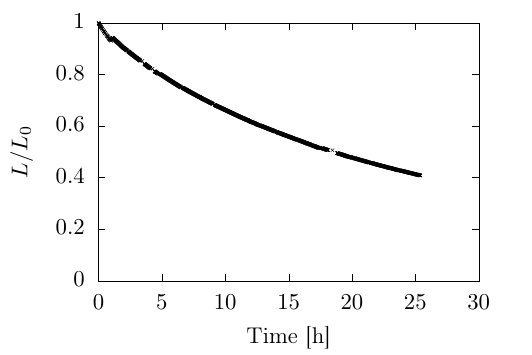}
    \caption{Luminosity data for the LHC Fill 5096 (2016) as found in the ``Massi'' files (left) and after data cleaning and normalisation (right).}
    \label{fig:dataclean}
\end{figure}

After this stage of fill selection, the raw luminosity data available in the LHC ``Massi'' files are still not suitable for applying the luminosity model, as they contain spurious data and the effect of the optimisation process known as luminosity scans\footnote{Given the tendency of the colliding beams to separate at the collision points due to ground motion of the insertion magnets, head-on collisions need to be re-established with regularity and this introduces transient behaviours in the luminosity evolution, which need to be filtered out.} that must be removed from each luminosity data set. An example of the raw luminosity data (left), together with the processed version (right), is shown in Fig.~\ref{fig:dataclean}. As a last step, the luminosity fill data are normalised as a function of its maximum value $L_0$, and the time is rescaled to be expressed in terms of LHC turns.

\begin{figure}
\centering
\includegraphics[width=.45\textwidth]{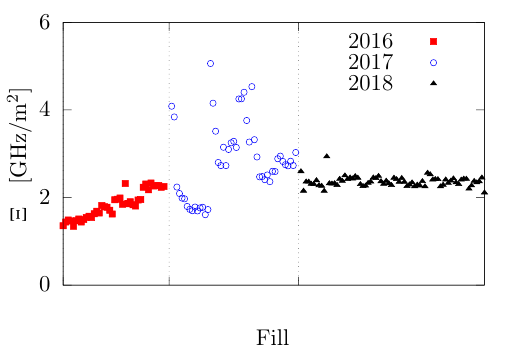}%
\includegraphics[width=.45\textwidth]{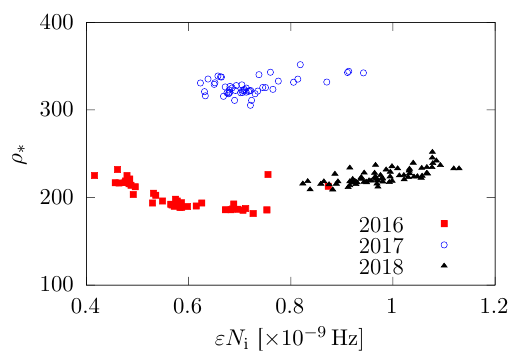}
\includegraphics[width=.45\textwidth]{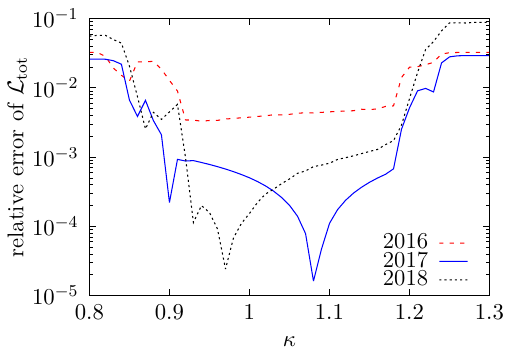}
\includegraphics[width=.45\textwidth]{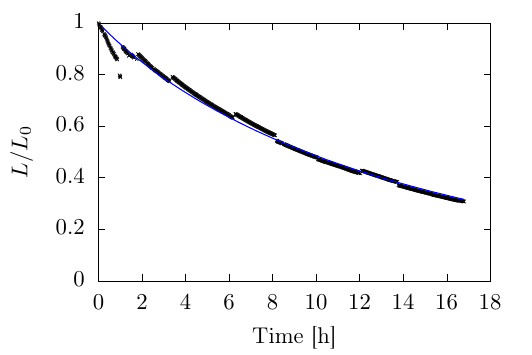}
    \caption{Top left: value of $\Xi=L_0/N_\text{i}^2$ as retrieved from the data of the fills chosen to represent the 2016, 2017 and 2018 runs. Top right: results of the fits of the model of Eq.~\ref{eq:lumimodel} for the selected fills of the three years in the $(\eps \nin, \rho_\ast)$ space, with fixed $\kappa$ ($\kappa_{2016}=\num{0.94}$, $\kappa_{2017}=\num{1.08}$, 
 $\kappa_{2018}=\num{0.97}$). Bottom left: relative difference in $\lt$ for each year using different fixed values of $\kappa$. The minima have been chosen as the constant $\kappa$ to use for the fits of each year. Bottom right: example of processed data of normalised luminosity as a function of time an the fit according to Eq.~\ref{eq:lumimodel} (blue line) for a single fill (\#6255) of the 2017 run.}
    \label{fig:fitresult}
\end{figure}

Fitting the model of Eq.~\eqref{eq:lumimodel} with its free parameters ($\rho_\ast$, $\varepsilon N_\text{i}$ and $\kappa$) is not straightforward, as they are highly correlated. Then, one needs to provide as much as possible physical information to guide the fit algorithm to find consistent values. First, the value of $\varepsilon N_\text{i}$ can, in principle, be retrieved from its definition. In fact, we have
\begin{equation}
    \varepsilon N_\text{i} = \sigma_\text{int} n_\text{c} \Xi N_\text{i} = \sigma_\text{int} n_\text{c} L_0/N_\text{i}\, ,
\end{equation}
and it should be possible to compute $\Xi$ as in Eq.~\eqref{eq:burnoff}, knowing the published optical parameters for each year of LHC operation and the initial beam intensity of each fill. However, a direct inspection of the results shown in Fig.~\ref{fig:fitresult} (top left) shows that the value of $\Xi=L_0/N_\text{i}^2$ appears to vary during each year, with the notable exception of 2018. Furthermore, it is worth mentioning that the values of the transverse and longitudinal emittances might vary from fill to fill and are hard to measure accurately. Therefore, we decided to determine $\varepsilon N_\text{i}$ as $\sigma_\text{int} n_\text{c} L_0$, and to allow the fitting algorithm to adjust its value in a window that ranges between $90\%$ and $110\%$ of the computed value.

The two remaining parameters of the luminosity model are the exponent $\kappa$ and the scaling factor $\rho_\ast$. According to Nekhoroshev's theorem~\cite{Nekhoroshev:1977aa}, $\kappa$ is only a function of the phase-space dimension, while $\rho_\ast$ is linked to the convergence radius of perturbative series, which in turn is connected to the strength of the nonlinearities present in the system. We then assume $\kappa$ to be constant during a year of operation (when the LHC ring parameters that determine the dynamical aperture are left unchanged), while we can assume that $\rho_\ast$ varies on a fill-to-fill basis, as the nonlinearities, e.g. due to the beam-beam effects, might vary in different fills. The best strategy is then to fit all the fills of a given year using the same value of $\kappa$ and choose the best one according to a specific metric: in our case, the fidelity to reproduce the integrated luminosity over a given year. We plot the result of this scan in Fig.~\ref{fig:fitresult} (bottom left), where we show for each value of $0.8<\kappa<1.3$ the absolute relative difference of the integrated luminosity obtained using the fitted model and the measured value. We see that for 2017 and 2018 we have a very clear optimal value of $\kappa$, given by the absolute minimum, while in 2016, although a minimum is visible for $\kappa=0.94$, the fitting algorithm finds a good, although never excellent, agreement for the interval $0.9 < \kappa <1.2$. This may be due to the fact that in the first year of operation of Run~2 the ring configuration could have been less stable than in the other years, thus affecting the relevant value of $\kappa$ for the model.

The results of the fit of $\rho_\ast$ with $\kappa$ kept constant at its optimal value are shown in the top-right plot of Fig.~\ref{fig:fitresult}, where we can observe the clustering of $\rho_\ast$ around similar values for each year. Finally, a typical result of the luminosity fit is shown in the bottom-right plot of the same figure. The discrete jumps of the luminosity data are generated by the beam adjustments aimed at optimising the collision conditions. In this study, we decided to fit the luminosity evolution by averaging these local variations, although it is envisaged to consider independently each of these sections in future studies.

An essential point is the stability of the fit parameters and in particular $\rho_\ast$ depending on the time span of the luminosity data in each fill used to perform the fit of the luminosity model. This analysis is important for two different aspects: first, it provides information on the physical significance of the model, as if the model parameters really reflect physical properties that should be constant during the fill, their estimation should rapidly converge to the true value; second, as we also mentioned above, the possibility of correctly guessing the fit parameters with limited data on the fill is necessary to estimate early on, i.e.\ during the evolution of a physics fill, the optimal fill time in the framework of an online optimisation procedure. 

\begin{figure}[htb]
\centering
\includegraphics[width=\textwidth]{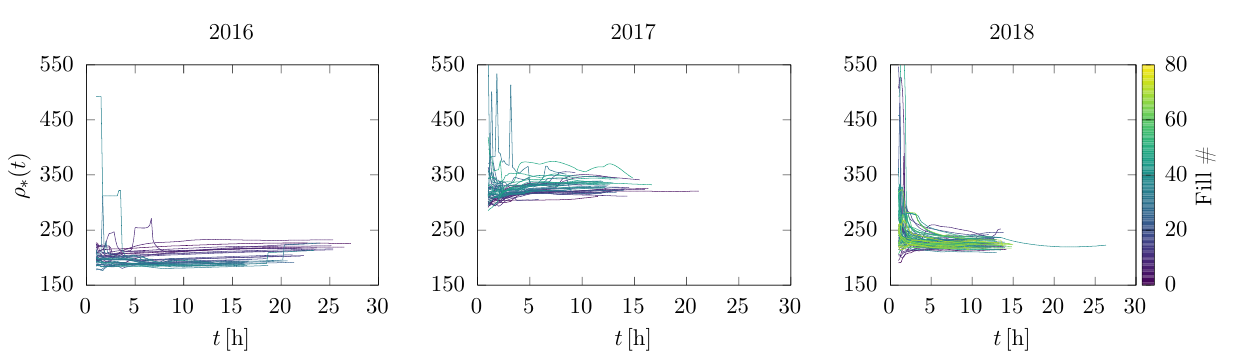}
\caption{Evolution of the fit parameter $\rho_\ast$ as a function of the time span parameter for each fill in the 2016, 2017 and 2018 pools having performed the fit according to Eq.~\eqref{eq:lumimodel} on the base only of data observed until time $t$. The colours identify the fills according to their ordinal position in the year's run.}
\label{fig:partialfit}
\end{figure}

Figure~\ref{fig:partialfit} shows the evolution of the fitted value of $\rho_\ast$ for each selected fill of Run~2 as a function of the time span in the physics fill. We see that most of the fills tend to rapidly, i.e.\ less than \SI{5}{h}, converge to the typical range range of $\rho_\ast$. The difficulty in estimating the right model parameters over a short time period is due to phenomena like that visible in the fill shown in the bottom-right plot of Fig.~\ref{fig:fitresult}, which presents some irregularities at the beginning, probably due to initial adjustments in the machine parameters. A slow drift of the value of $\rho_\ast$ with time is observable since the fit is also simultaneously adjusting the value of $\eps N_\text{i}$ in its freedom window. Ultimately, the drift is the result of a complex model like Eq.~\ref{eq:lumimodel} where the parameters are highly correlated and, also in this case, future research will be devoted to make the fit more robust and efficient.

\subsection{Possible gain in integrated luminosity from optimisation of the fill time} \label{sec:post}
The first aspect that can be considered when trying to apply the proposed optimisation strategy to the case of the LHC Run~2 is whether there is anything to gain in changing the actual value of $\htfc$. This means to perform an \textit{a posteriori} optimisation of each the fill times of each year. In fact, each of these fills have been kept in the actual run for a given duration, but we could consider re-distributing the fill times among the fills, keeping constant the total time and the number of fills, verifying whether the integrated luminosity can be increased.

Let $F_y$ be the set of fills selected for year $y$ and $\nf$ be the corresponding number of fills. Each element of $F_y$ can be seen as a tuple $(N_{\text{i},j}, \varepsilon_{j}, \rho_{\ast, j}, \kappa_j, \ttaj, t_{\text{fill},j})$ where $\nin$, $\varepsilon$, $\kappa$ and $\rho_\ast$ are the fit parameters of the luminosity model, $\ttaj$ the turnaround time before the fill, and $t$ the actual fill length. If we define the integrated luminosity of the year $y$ for the vector of times $\mathbf{t}_\mathrm{act}=(t_{\text{fill},1}, \cdots, t_{\text{fill},\nf})$ as
\begin{equation}
    \mathcal{L}_y(\mathbf{t}_\text{act}) = \sum_{j\in F_y} \int_0^{\tfj} \dd t\, L(t; N_{\text{i},j}, \varepsilon_j, \kappa_j, \rho_{\ast, j}) \, , 
\end{equation}    
our goal consists in finding values $\htfj$ such that $\mathcal{L}_y(\hat{\mathbf{t}})$
is maximal, with the constraint
\begin{equation}
T_y = \sum_{j\in F_y} \htfj = \sum_{j\in F_y} \tfj \, .
\end{equation}

Defining $\mathcal{L}_{\text{post},y}$ the maximum value obtained with this optimisation \textit{a posteriori}, i.e.\ $\mathcal{L}_y(\hat{\mathbf{t}})$, and $\mathcal{L}_{\text{act},y}$ as the actual value, i.e.\ $\mathcal{L}_y(\mathbf{t}_\text{act})$, we obtain a possible increase in integrated luminosity in the range of $2\%$-$3\%$ for the selected fills of the three years. The detailed results are shown in the first four columns of Table~\ref{tab:opt_results-stat}.

\begin{figure}[htb]
\centering
\hspace*{-5.5truemm}
\includegraphics[trim=0truemm 0truemm 3truemm 0truemm, width=.36\textwidth,clip=]{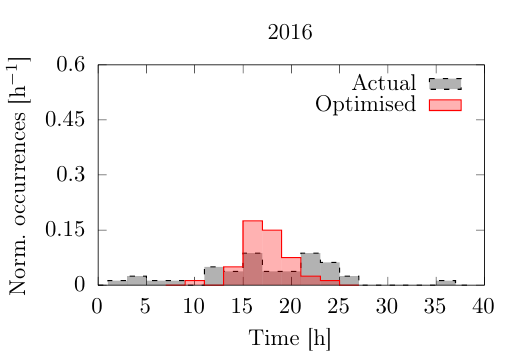}
\hspace*{-3.0truemm}
\includegraphics[trim=7truemm 0truemm 3truemm 0truemm, width=.33\textwidth,clip=]{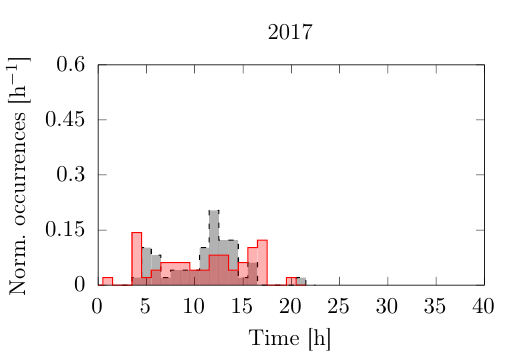}
\includegraphics[trim=7truemm 0truemm 3truemm 0truemm, width=.33\textwidth,clip=]{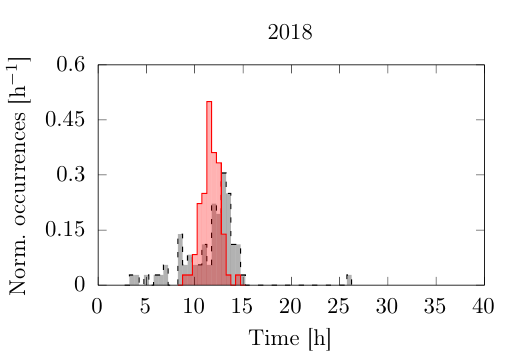}
\includegraphics[width=\textwidth]{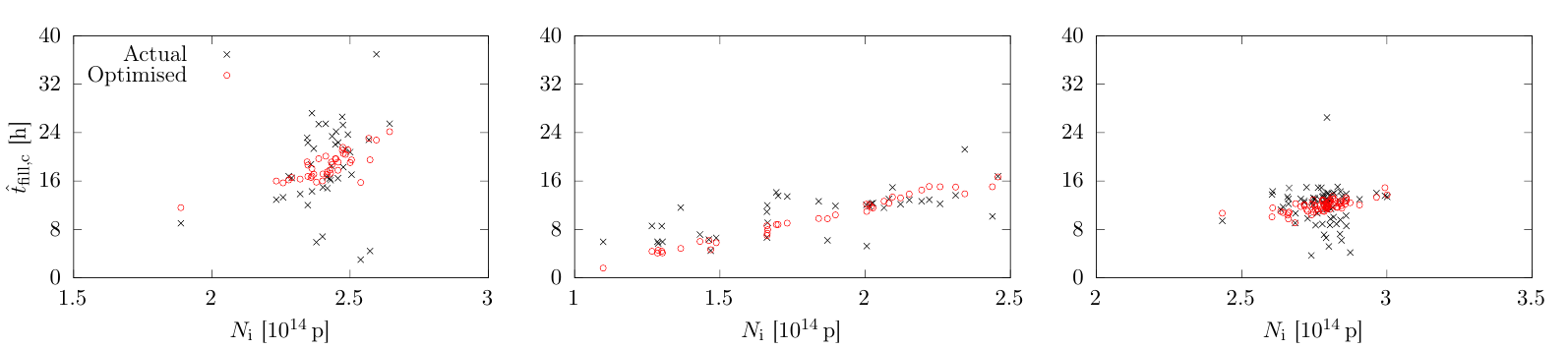}
\includegraphics[width=\textwidth]{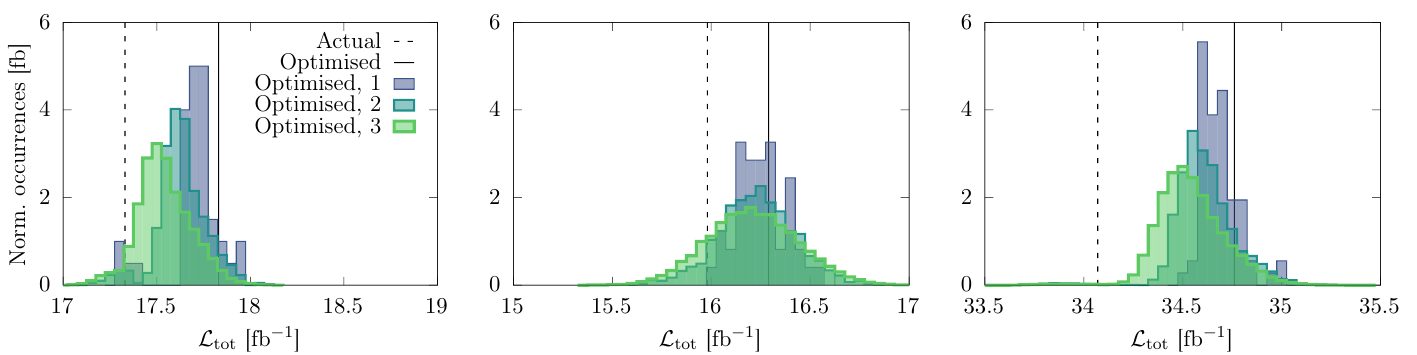}
\caption{Top row: Histograms representing the actual and optimised distribution of the fill lengths in the selections of the fills for 2016, 2017 and 2018. Centre row: Comparison of the actual (black crosses) and optimised (red circles) fill lengths as function of the initial beam population $\nin$. Bottom row: Distributions of possible integrated luminosities when optimising the fill length with the \textit{a posteriori} approach. The optimisation is performed also removing all possible combinations of 1 (grey), 2 (blue) or 3 (green) fills from the selections, although keeping the total time for physics constant. The solid black line represents the \textit{a posteriori} optimisation of the full set of fills, while the dashed line is the actual integrated luminosity of the set of fills.}
\label{fig:compare_apost}
\end{figure}

The comparison between the histograms of the optimal fill times and the original ones is shown in Fig.~\ref{fig:compare_apost} (top row). In general, the optimiser has a strong tendency to concentrate the fill times around an optimal duration. In particular, for 2016 and 2018, where the observed initial distribution of fill times spans a short interval, the optimised time distribution shrinks to something close to a central value. However, the situation in 2017 appears to be more complex and the origin of this behaviour is found when we plot the correlation between the length of the fill against $\nin$, as in Fig.~\ref{fig:compare_apost} (centre row). It becomes clear that in 2017 the fills under consideration span a much wider interval of $\nin$ (approximately $\SI{2e14}{\proton}$ intensity variation, to be compared with approximately $\SI{4e13}{\proton}$ intensity range in which most of the fills from 2016 and 2018 are contained) and a clear correlation between $\htfc$ and $\nin$.

The proposed \textit{a posteriori} optimisation provides the best possible performance, in terms of integrated luminosity, having complete information on the fills in the pool. However, it makes the implicit assumption that the number of fills is already optimal. This aspect can be verified and studied by transforming the selected sets of fills. The first transformation consists of removing one fill from the initial set in all possible combinations. Then, the \textit{a posteriori} optimisation is applied to all groups of fills with one less fill than the initial situation. A second transformation is applied by creating a new ensemble of fills in which each possible combination of pairs of fills is removed. In this case also the \textit{a posteriori} optimisation is applied. The last case corresponds to the case in which all possible compilations of three fills are removed. Note that for all three types of transformation, the constraint to the \textit{a posteriori} optimisation is given by the same total time for physics.  

The histograms representing the distribution of the total integrated luminosity achievable under these conditions are represented in Fig.~\ref{fig:compare_apost} (bottom row). We observe, in general, that removing fills and redistributing their time does not perform, on average, better than the \textit{a posteriori} optimisation of the entire fill set (with the exception of 2017 where the peak of the three distributions is very close to the value obtained without removals). The histograms also show that there is a certain fraction of distributions that are beyond the total integrated luminosity achieved with the \textit{a posteriori} optimisation. The results of this analysis are summarised in Table~\ref{tab:opt_results-stat}.

Overall, this suggests that reducing the number of fills is not the right direction to follow to obtain better results. We should rather try to increase the fill pool, and for this we switch to the more realistic optimisation, which is the focus of this paper, using our Monte Carlo generator to augment with plausible data each year's fills.

\begin{table}
\centering
\caption{Summary of the integrated luminosity for the various cases of optimisation strategy. For the distributions shown in Fig.~\ref{fig:compare_apost} (bottom row) the average, standard deviation and probability of observing an integrated luminosity larger than $\mathcal{L}_\text{post}$ are shown.}
\begin{tabular}{c|c|c|c|c|c|c|c|c|c}
\hline 
  &   &                             &                              &  \multicolumn{2}{c|}{1 fill removed}  & \multicolumn{2}{c|}{2 fills removed}  & \multicolumn{2}{c}{3 fills removed}  \\
Year &  T & $\mathcal{L}_{\text{act}}$ &  $\mathcal{L}_{\text{post}}$ &  $\mathcal{L}$  & $\mathcal{L}>\mathcal{L}_{\text{post}}$   & $\mathcal{L}$  & $\mathcal{L}>\mathcal{L}_{\text{post}}$   & $\mathcal{L}$ & $\mathcal{L}>\mathcal{L}_{\text{post}}$    \\
     & [\si{\hour}]& [\si{\femto\barn^{-1}}] & [\si{\femto\barn^{-1}}]& $[\si{\femto\barn^{-1}}]$ & [$\%$] & $[\si{\femto\barn^{-1}}]$ &  [$\%$] &  $[\si{\femto\barn^{-1}}]$ &  [$\%$]  \\
\hline
2016 & \num{737} & \num{17.33} & \num{17.83}  & \num{17.72 \pm 0.14} & 15.0 & \num{17.66 \pm 0.13} & 9.7 & \num{17.57 \pm 0.14} & 4.0 \\
2017 & \num{559} & \num{15.98} & \num{16.29} & \num{16.29 \pm 0.13} & 44.9 & \num{16.26 \pm 0.19} & 42.0 & \num{16.24 \pm 0.23} & 40.2 \\
2018 & \num{857} & \num{34.07} & \num{34.76}  & \num{34.71 \pm 0.10} & 20.8 & \num{34.64 \pm 0.15} & 17.5 & \num{34.57 \pm 0.17} & 13.0 \\
\hline
\end{tabular}
\label{tab:opt_results-stat}
\end{table}
\subsection{Simulation of possible Run~2 performance using integrated luminosity optimisation}
The optimisation approach discussed in Section~\ref{sec:post} has an intrinsic limitation as it can only be performed \textit{a posteriori}, i.e. considering all the fills of a given year simultaneously. Furthermore, it only allows us to act on the duration of existing fills, redistributing time between them, and it was shown that there is no guarantee that the actual number of fills in each year is optimal. In fact, the total integrated luminosity could be increased cutting shorter all fills and leaving some time for additional ones. To assess the possible increase in integrated luminosity in this situation, we can run the proposed optimisation algorithm, which considers the fills sequentially, over the actual fills of a given year, and estimate the possible attainable luminosity considering a total time equal to the original one either by projecting the luminosity obtained during the actual number of fills to the total time for physics, or use a Monte Carlo generator to create additional fills until the yearly time is reached.

\begin{figure}[htb]
\centering
\includegraphics[width=\textwidth]{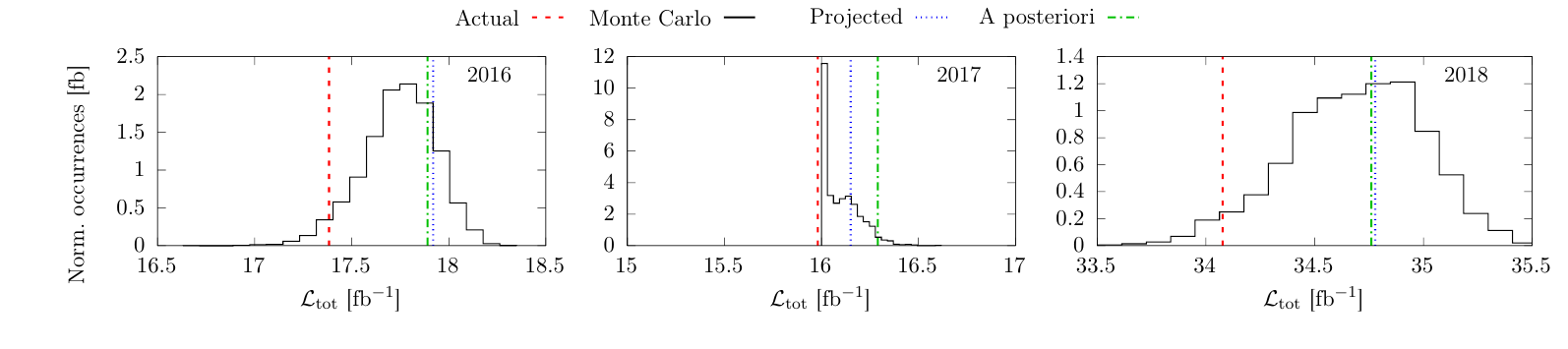}
\caption{Values of the integrated luminosity and possible optimisations according to different approaches for the 2016, 2017 and 2018 fill sets. The black distribution represents the data obtained from \num{1e4} runs of the Monte Carlo generator, while the other lines represent the actual luminosity, the projected value based on the online optimisation of the real fills and the value obtained by the \textit{a posteriori} optimisation.}
\label{fig:compare_mixed}
\end{figure}

The results for each year are shown in Fig.~\ref{fig:compare_mixed}. The histogram represents the distribution of the total integrated luminosity achieved when fills generated by a Monte Carlo method are added to the real ones until the total time for physics is attained. The lines instead show the values of the actual luminosity as well as the result of the \textit{a posteriori} optimisation. The value of the projected luminosity, obtained as
\begin{equation}
\mathcal{L}_{\text{proj},y} = \frac{\mathcal{L}_y(\mathbf{t}_\text{online})}{T_{\text{ta}, y} + \sum_{j\in F_y} t_{\text{online},j}} (T_y + T_{\text{ta},y}) \, ,  
\end{equation}
where $\mathbf{t}_\text{online}$ is the vector of the fill times obtained by the proposed optimisation strategy and $T_{\text{ta},y}$ is the total turnaround time for the fills in $F_y$, is also shown.

For 2016 and 2018, very close values of $\mathcal{L}_\text{proj}$ and $\mathcal{L}_\text{post}$ are found. However, this observation is not generic. In fact, while the \textit{a posteriori} optimisation has the advantage of perfect knowledge of the fills in the pool, the online algorithm updates its best estimate of future fills according to the statistics of the previous ones. On the other hand, the proposed optimisation approach can also act by increasing the number of fills. Furthermore, we can explain the lower efficiency of the optimisation strategy in the 2017 case with respect to the \textit{a posteriori} optimisation in the greater variability of $\nin$ during that year, making it more difficult for the online algorithm to determine a reliable estimate of the parameters of the future fills. 

It should also be noted that for the new fills generated at the end of the year, we used the value of $\Xi$ calculated from the values reported in Table~\ref{tab:datarun2}. However, the observations of Fig.~\ref{fig:fitresult} (top left) are not always consistent with a constant value, and this could explain why the projected values of the total integrated luminosity appear to be better than the average of the Monte Carlo distribution shown. All simulation results are summarised in Table~\ref{tab:opt_results}.

\begin{table}
\centering
\caption{Summary of the actual integrated luminosity and of possible optimisations strategies for the 2016, 2017, and 2018 fill sets. The quoted increase in integrated luminosity is computed with respect to the value of $\mathcal{L}_{\text{act}}$. For the case of $\mathcal{L}_{\text{Monte Carlo}}$ the average and the standard deviation are quoted. The values in brackets represent the increase w.r.t.\ $\mathcal{L}_\text{act}$.}
\begin{tabular}{c|c|c|c|c|c}
\hline 
Year & T & $\mathcal{L}_{\text{act}}$ &  $\mathcal{L}_{\text{post}}$ &  $\mathcal{L}_{\text{proj}}$ & $\mathcal{L}_{\text{Monte Carlo}}$ \\
     & [\si{\hour}] & [\si{\femto\barn^{-1}}] & [\si{\femto\barn^{-1}}]& [\si{\femto\barn^{-1}}] & [\si{\femto\barn^{-1}}] \\
\hline
2016 & 737 & \num{17.33} & \num{17.83} ($+2.9\%$) & \num{17.92} ($+3.4\%$) & \num{17.74 \pm 0.19} \\
2017 & 559 & \num{15.98} & \num{16.29} ($+2.0\%$) & \num{16.15} ($+1.1\%$) & \num{16.10 \pm 0.09} \\
2018 & 857 & \num{34.07} & \num{34.76} ($+2.0\%$) & \num{34.78} ($+2.1\%$) & \num{34.69 \pm 0.31} \\
\hline
\end{tabular}
\label{tab:opt_results}
\end{table}

The various approaches presented and applied to the Run~2 data consistently confirm the possibility of improving the total integrated luminosity by a non-negligible amount.
\section{Conclusions and outlook} \label{sec:conc}
This paper presents and thoroughly examines two methods for determining an optimal length of a fill for physics for a circular collider. Monte Carlo simulations, meticulously aligned with the typical distributions found in the LHC Run~2 datasets, reveal that the proposed optimisation strategy significantly enhances integrated luminosity. Motivated by these encouraging outcomes, the LHC Run~2 physics runs have been reevaluated and rerun, in numerical simulations, using the optimised method. These assessments demonstrate that by using the suggested strategy for determining an optimised fill length, a luminosity increase of a few percent is within reach.It is important to emphasise that this advancement can be achieved without introducing any additional specialised hardware. Instead, it can be accomplished through an optimised approach for determining the duration of each fill, using software tools.

These encouraging results serve as the foundation for future research, which will concentrate on applying these methods to the scenario of levelled luminosity, a standard technique for the LHC Run~3 and a key aspect of the upgraded LHC ring. Future investigations will also explore the feasibility of delivering an efficient and reliable estimate of the luminosity for upcoming fills, leveraging data from previous fills. In addition, a comprehensive model for the evolution of luminosity will be developed to ensure a reliable estimate of the fill duration, assuming real-time luminosity measurement. The implications of measurement errors on luminosity will be evaluated to gauge the sensitivity of the proposed method against various characteristics of luminosity measurement.
\section*{Acknowledgements}
The authors thank A. Brizzi and D. Mirarchi for several fruitful discussions. We also extend our gratitude to the conveners and members of the ATLAS Luminosity Working Group for their valuable discussions and feedback.
\section*{Data availability}
Data sets generated during the current study are available from the corresponding author on reasonable request. 

\clearpage
\appendix
\section{Evolution of beam intensity and its link with the DA} \label{sec:app}
\subsection{The case of a Gaussian beam distribution}
In this section, we review the link between DA and the intensity evolution proposed in~\cite{da_and_losses}. In the case considered here, the assumption is that the DA is computed in the action space, i.e.\ using $J_x$ and $J_y$~\cite{titze:2021}. The starting point is a Gaussian distribution in the $(u,p_u)$ variables ($u=x, y$), namely,
\begin{equation}
\rho(u,p_u; \epsilon) = \frac{1}{2 \pi \epsilon} \Exp{-\left(\frac{u^2}{2 \epsilon} + \frac{p_u^2}{2 \epsilon} \right)}
\label{eq:hat_rho}
\end{equation}
so that the actual transverse beam distribution reads
\begin{equation}
f(x,p_x, y, p_y) = N_\mathrm{p} \rho(x,p_x; \epsilon_{x}) \rho(y, p_y; \epsilon_{y}) \, ,
\end{equation}
where $N_\mathrm{p}$ represents the total beam intensity. 

Using $x^2 + p_x^2 = 2 J_x$ and $y^2 + p_y^2 = 2 J_y$, we can express $f$ in terms of angle-action variables:
\begin{equation}
f(J_x, J_y, \varphi_x, \varphi_y) = \frac{N_\mathrm{p}}{4 \pi^2 \epsilon_{x} \epsilon_{y}} \Exp{-\left(\frac{J_x}{\epsilon_{x}} + \frac{J_y}{\epsilon_{y}}\right)}\, ,
\end{equation}
and 
\begin{equation}
\hat f(J_x, J_y) = \int_0^{2 \pi} \int_0^{2 \pi} f(J_x, J_y, \varphi_x, \varphi_y) \dd \varphi_x \dd \varphi_y = \frac{N_\mathrm{p}}{\epsilon_{x} \epsilon_{y}} \Exp{-\left(\frac{J_x}{\epsilon_{x}}+ \frac{J_y}{\epsilon_{y}}\right)}\, .
\end{equation}

Note that the usual relation
\begin{equation}
\langle J_u \rangle = \int_0^\infty \int_0^\infty J_u \hat f(J_x, J_y) \dd J_x \dd J_y = \epsilon_u \, ,
\end{equation}
where $u=x, y$, is fulfilled. 

We introduce the coordinates $r \in [0, \infty[$ and $\theta \in [0, \pi/2]$ as follows:
\begin{subequations}
\begin{align}
\sqrt{J_x} = \sqrt{\epsilon_x} r \cos \theta \, , \\
\sqrt{J_y} = \sqrt{\epsilon_y} r \sin \theta \, ,
\end{align}
\end{subequations}
and we define the following quantity
\begin{equation}
r_\mathrm{max}(\theta)=\frac{R \sqrt{\epsilon_x \epsilon_y}}{\sqrt{\epsilon_y\cos^2\theta + \epsilon_x \sin^2\theta}} \qquad R \in [0, \infty[ \, .
\label{eq:rmax}
\end{equation}

We are interested in the fraction of particles contained within a specific region given by $r \leq r_\mathrm{max}(\theta) , \theta \in [0, \pi/2 ] $. This corresponds to computing the surviving particles $S$, i.e.\ those particles that are located inside the DA, which is given by
\begin{equation}
S(R) = \int_0^\infty \int_0^\infty \Theta_R(r, \theta) \hat f(J_x, J_y) \dd J_x \dd J_y , \hspace{1cm}
\Theta_R(r, \theta) = \left\{\begin{array}{cl} 1 & \text{for $r \leq r_\mathrm{max}(\theta)$} \\ 0 & \text{for $r > r_\mathrm{max}(\theta)$ \, .} \end{array} \right .
\label{eq:rd}
\end{equation}

From the definition, we have
\begin{equation}
dJ_x dJ_y = 4 \epsilon_x \epsilon_y r^3 \cos \theta \sin \theta dr d\theta \, , \label{eq:djxdjy}
\end{equation}
and 
\begin{equation}
\frac{J_x}{\epsilon_x} + \frac{J_y}{\epsilon_y} = r^2 \, .
\label{eq:jxjy1}
\end{equation}

Inserting Eqs. \eqref{eq:djxdjy}, and \eqref{eq:jxjy1} in Eq. \eqref{eq:rd} yields:
\begin{equation}
S(R) = 4 N_\mathrm{p} \int_0^{\pi/2} \dd\theta \int_0^{r_\mathrm{max}(\theta)}  \dd r\, 
 \Exp{-r^2} r^3 \cos \theta \sin \theta 
 \label{eq:losses0}
\end{equation}
and the integral in Eq.~\eqref{eq:losses0} becomes:
\begin{equation}
\begin{split} 
S(R) &=4N_\text{p} \int_0^{\pi/2}\dd\theta\,\cos\theta\sin\theta \int_0^{r_\mathrm{max}(\theta)}\dd r\, r^3 e^{-r^2} \\ 
&= 2N_\text{p} \int_0^{\pi/2} \dd\theta\,\sin\theta\cos\theta \qty[1-(1+r^2_\mathrm{max}(\theta))e^{-r^2_\mathrm{max}(\theta)}] \, .
\end{split}
\end{equation}

With the substitution $w=\epsilon_y\cos^2\theta + \epsilon_x \sin^2\theta$, we get $\dd w = -2(\epsilon_y-\epsilon_x)\cos\theta\sin\theta\dd\theta$ and we can rewrite the integral as
\begin{equation}
\begin{split}
    S(R) & =N_\text{p}\qty[ 1 - \int_{\epsilon_y}^{\epsilon_x} \frac{\dd w}{\epsilon_x - \epsilon_y} \qty(1+\frac{R^2 \epsilon_x \epsilon_y}{w})\exp(-\frac{R^2 \epsilon_x \epsilon_y}{w})] \\
    & = N_\text{p}\qty[1 - \frac{\epsilon_x e^{-R^2 \epsilon_y} - \epsilon_y e^{-R^2 \epsilon_x}}{\epsilon_x - \epsilon_y} ]
\end{split}
\label{eq:subsint}
\end{equation}

Note that considering $\epsilon_y=\epsilon_x+\Delta$, one obtains that
\begin{equation} S(R)=N_\text{p}\qty[1 - \qty(1+\epsilon_x R^2-\epsilon_x \frac{R^4}{2} \Delta )e^{-R^2\epsilon_x}] \, . 
\end{equation} 
and the limit $\epsilon_y\to\epsilon_x=\epsilon$ or $\Delta \to 0$ gives
\begin{equation} S(R)=N_\text{p}\qty[1 - (1+\epsilon R^2)e^{-R^2\epsilon}] \, . 
\end{equation} 

All these relationships represent the number of charged particles in the region $r \le r_\mathrm{max}(\theta)$. However, $R=D(N)$, where $D(N)$ is the DA value that varies with the number of turns $N$. Hence, 
\begin{equation}
S(N) = N_\mathrm{p} \qty[1 - \frac{\epsilon_x e^{-D(N)^2 \epsilon_y} - \epsilon_y e^{-D(N)^2 \epsilon_x}}{\epsilon_y - \epsilon_x} ] \, ,
\label{eq:losses1}
\end{equation}
or
\begin{equation} 
S(R)=N_\text{p}\qty[1 - \qty(1+\epsilon_x D(N)^2-\epsilon_x \frac{D(N)^4}{2} \Delta )e^{-D(N)^2\epsilon_x}] 
\end{equation} 
for the case of emittances with a small difference $\Delta$ or 
\begin{equation} 
S(R)=N_\text{p}\qty[1 - (1+\epsilon D(N)^2)e^{-D(N)^2\epsilon}] 
\end{equation} 
in the case of equal emittances.

\subsection{The case of a non-Gaussian beam distribution}\label{sec:qauss}
The use of so-called $q$-Gaussian distributions is expanding in accelerator physics~\cite{PhysRevAccelBeams.23.101004} as it provides a good description of transverse beam distributions with heavy tails, i.e.\ tails heavier than those of a Gaussian distribution. Therefore, we consider a luminosity model based on $q$-Gaussian distributions. 

The first step consists of defining the transverse distribution. To this end, we introduce a $4$D $q-$Gaussian distribution (with $1< q < 2$) to model the presence of heavy tails in the transverse distribution. In action-angle variables, the distribution reads
\begin{equation} \rho(J_x,J_y,\phi_x,\phi_y) = \frac{\lambda_x\lambda_y}{\pi^2}(2q-3)(q-2)e_q(-2\lambda_x J_x-2\lambda_y J_y)\,,\label{eq:qgauss}\end{equation}
where
\begin{equation}e_q(x) = [1+(1-q)x]^{1/(1-q)},\qquad e_q(x)\to e^x\,\text{for }q\to 1.\end{equation}

We can verify the normalisation of Eq.~\eqref{eq:qgauss} by defining $B_x=\lambda_x(1-q)$, $B_y=\lambda_y(1-q)$ and $\alpha=1/(1-q)$, and we obtain
\begin{equation}
    \begin{split} 
    \int_0^\infty \dd J_x \int_0^\infty \dd J_y \int_0^{2\pi} \dd\phi_x \int_0^{2\pi} & \dd \phi_y\, \rho(J_x,J_y,\phi_x,\phi_y) = \\ 
    = & 4\lambda_x\lambda_y(2q-3)(q-2) \int_0^\infty \dd J_y \int_0^\infty \dd J_x\, (1+2B_x J_x + 2B_y J_y)^\alpha \\ 
    = & -\frac{2\lambda_x\lambda_y(2q-3)(q-2)}{B_x (\alpha +1)} \int_0^\infty \dd J_y\, (1+B_y J_y)^{\alpha+1} \\
    = & \frac{\lambda_x\lambda_y(2q-3)(q-2)}{B_x B_y (\alpha+1)(\alpha+2)} \\ 
    = & 1 \, .
    \end{split}
\end{equation}

Then, we proceed as in the case of a Gaussian distribution, namely, we introduce the $(r,\theta)$ coordinates as
\begin{subequations}
\begin{align}
\sqrt{J_x} = \frac{1}{\sqrt{2\lambda_x}} r \cos \theta \, , \\
\sqrt{J_y} = \frac{1}{\sqrt{2\lambda_y}} r \sin \theta \, ,
\end{align}
\end{subequations}
and we compute
\begin{equation}S(R) = 4N_\mathrm{p}\int_0^\infty \dd J_x \int_0^\infty \dd J_y \Theta_R(r,\theta) \rho(J_x,J_y)\,. \end{equation}
where $\Theta_R(r,\theta)$ is defined in a way similar to Eq.~\eqref{eq:rd}, with
\begin{equation}
r_\mathrm{max}(\theta)=\frac{R}{\sqrt{\lambda_x \cos^2 \theta + \lambda_y \sin^2 \theta}}
\qquad R \in [0, \infty[ \, .
\label{eq:rmax1}
\end{equation}

The change of variables gives the Jacobian $r^3\cos\theta\sin\theta/\lambda_x\lambda_y$ and the integral becomes
\begin{equation}
\begin{split}
\frac{S(R)}{N_\mathrm{p}} &= 4(2q-3)(q-2)\int_0^{\pi/2}\dd \theta\,\sin\theta\cos\theta \int_0^{r_\text{max}(\theta)} \dd r\, r^3 [1-(1-q)r^2]^{1/(1-q)}\\
&= 2\int_0^{\pi/2}\dd \theta\,\sin\theta\cos\theta \qty{1 - [1 + r_\text{max}(\theta)^2 - r_\text{max}(\theta)^4(q-1)(q-2)]e_q(-r_\text{max}(\theta))^2)}\,.
\end{split}
\end{equation}
Introducing the variable $w$ as in Eq.~\eqref{eq:subsint}, we rewrite the integral as
\begin{equation}
\begin{split}
\frac{S(R)}{N_\mathrm{p}} &= 1- 2\int_{1/2\lambda_x}^{1/2\lambda_y} \dd w \frac{(\lambda_y-\lambda_x)}{\lambda_x\lambda_y} \qty[1+\frac{R^2}{4\lambda_x\lambda_y w} - \frac{R^4}{16\lambda_x^2\lambda_y^2 w^2}(q-1)(q-2)] \times \\
& \times \qty[1-(1-q)\frac{R^2}{4\lambda_x\lambda_y w}]^{\frac{1}{1-q}}\\
&=1-\frac{\lambda_y-\lambda_x}{\lambda_x\lambda_y}\left\{\frac{e_q(-R^2/2\lambda_x)}{\lambda_y} \qty[1+\frac{R^2}{\lambda_x}(q-1)+\frac{R^4}{4\lambda^2_x}(q-1)(q-2)]+\right. \\
&\qquad\qquad\qquad - \left. 
\frac{e_q(-R^2/2\lambda_y)}{\lambda_x} \qty[1+\frac{R^2}{\lambda_y}(q-1)+\frac{R^4}{4\lambda^2_y}(q-1)(q-2)]\right\}\,.
\end{split}
\end{equation}

Note that we retrieve the expression for the Gaussian case for $q\to 1$, when also $\lambda_{x,y}\to 1/2\epsilon_{x,y}$. If, on the other hand, we have $\lambda_y=\lambda_x$, the expression reduces to
\begin{equation}
    S(R) = N_\text{p} \qty{1 - \qty[1 + \frac{R^2}{2\lambda_x} - \frac{R^4}{4\lambda_x^2}(q-1)(q-2)]}e_q\qty(-\frac{R^2}{2\lambda_x})\,.    
\end{equation}

Of course, in the previous two expressions, we can replace $R$ with $D(N)$ to link the concept of surviving particles with that of DA.

\clearpage
\bibliographystyle{unsrt}
\bibliography{mybibliography,extrabib}
\end{document}